\documentclass[fleqn,usenatbib,useAMS]{mnras}

\usepackage{xcolor}

\usepackage{graphicx, float}	
\usepackage{amsmath}	
\usepackage{amssymb}	
\usepackage{multicol}        
\usepackage{bm}		
\usepackage{pdflscape}	
\usepackage{lipsum}
\usepackage{tipa}

\usepackage[T1]{fontenc}
\usepackage{ae,aecompl}

\usepackage{newtxtext,newtxmath}

\title[Merger Observability with IllustrisTNG]{The limitations (and potential) of non-parametric morphology statistics for post-merger identification}

\author[S. Wilkinson et al.]{Scott Wilkinson$^{1}$\thanks{Contact e-mail: \href{mailto:swilkinson@uvic.ca}{swilkinson@uvic.ca}}, Sara L. Ellison$^{1}$, Connor Bottrell$^{2,3,4}$, Robert W. Bickley$^{1}$, Shoshannah Byrne-Mamahit$^{1}$,\newauthor Leonardo Ferreira$^{1}$, and David R. Patton$^{5}$
\\
$^{1}$Department of Physics and Astronomy, University of Victoria, Victoria, British Columbia V8P 1A1, Canada\\
$^{2}$Kavli Institute for the Physics and Mathematics of the Universe (WPI), UTIAS, University of Tokyo, Kashiwa, Chiba 277-8583, Japan\\
$^{3}$International Centre for Radio Astronomy Research, University of Western Australia, 35 Stirling Hwy, Crawley, WA 6009, Australia\\
$^{4}$Center for Data-Driven Discovery, Kavli IPMU (WPI), UTIAS, The University of Tokyo, Kashiwa, Chiba 277-8583, Japan\\
$^{5}$Department of Physics and Astronomy, Trent University, 1600 West Bank Drive, Peterborough, ON K9L 0G2, Canada
}

\date{September 29th, 2022}

\pubyear{2022}

\begin{document}
\label{firstpage}
\pagerange{\pageref{firstpage}--\pageref{lastpage}}
\maketitle

\begin{abstract}
Non-parametric morphology statistics have been used for decades to classify galaxies into morphological types and identify mergers in an automated way. In this work, we assess how reliably we can identify galaxy post-mergers with non-parametric morphology statistics. Low-redshift ($z\lesssim0.2$), recent ($t_\text{post-merger} \lesssim 200$ Myr), and isolated ($r > 100$ kpc) post-merger galaxies are drawn from the IllustrisTNG100-1 cosmological simulation. Synthetic r-band images of the mergers are generated with SKIRT9 and degraded to various image qualities, adding observational effects such as sky noise and atmospheric blurring. We find that even in perfect quality imaging, the individual non-parametric morphology statistics fail to recover more than 55\% of the post-mergers, and that this number decreases precipitously with worsening image qualities. The realistic distributions of galaxy properties in IllustrisTNG allow us to show that merger samples assembled using individual morphology statistics are biased towards low mass, high gas fraction, and high mass ratio. However, combining all of the morphology statistics together using either a linear discriminant analysis or random forest algorithm increases the completeness and purity of the identified merger samples and mitigates bias with various galaxy properties. For example, we show that in imaging similar to that of the 10-year depth of the Legacy Survey of Space and Time (LSST), a random forest can identify 89\% of mergers with a false positive rate of 17\%. Finally, we conduct a detailed study of the effect of viewing angle on merger observability and find that there may be an upper limit to merger recovery due to the orientation of merger features with respect to the observer.
\end{abstract}

\begin{keywords}
galaxies:interaction, galaxies: evolution, galaxies: structure
\end{keywords}

\section{Introduction}
\label{Intro}


Galaxy mergers can drastically impact galaxy properties. Observations are consistent with a general merger sequence in which the initial pair phase of a galaxy interaction, gravitational torques distort the distribution of stars throughout the galaxy \citep{dePropris07, Casteels14, Patton16} and cause gas to lose angular momentum and fall inward to the centre of galaxies triggering central starbursts \citep{Barton00, Lambas03, Ellison08, Woods10, Patton13, Knapen15}, decreasing central gas-phase metallicity \citep{Kewley06, Scudder12}, and triggering active galactic nuclei \citep[AGN;][]{Ellison11,Satyapal14,Goulding18}. After coalescence of the merging galaxies, there remains an enhancement in star-formation activity \citep{Bridge10,Pearson19, Bickley22, Shah22}, as well as a suppression in metallicity \citep{Ellison13, Thorp19} and an increase in AGN activity \citep{Ellison19, Bickley23, Li23a}. As the central starburst fades, post-mergers have been shown to exhibit an enhancement in the frequency of post-starburst (PSB) spectral features indicating a complete transition from star-forming systems to quiescence \citep{Ellison22, Li23}. Thus, the global and substantial impact that galaxy interactions impart upon their subjects endues tremendous importance to the effort of assessing the presence of a recent or ongoing merger for many studies throughout the field of galaxy evolution.


Fortunately, theory and numerical simulations demonstrate that galaxy interactions and mergers produce disruptions and deformations in the stellar distributions of the affected galaxy \citep{Toomre72, Barnes92}. Features such as shells \citep{Quinn84, Barnes92}, tidal streams \citep{Negroponte83, Amorisco15}, and warped or asymmetric isophotes \citep{Naab03} are all unique indicators of a recent galaxy interaction. The disrupted stellar orbits can be traced with rest-frame-optical imaging, giving observers a clear way to identify recent mergers and interactions and subsequently study their properties or incidence rates. Thus many previous works have used rest-frame-optical imaging to identify recent mergers for the purpose of studying the mergers themselves \citep[e.g.][]{Bickley22, Bickley23, Ellison19, Ellison22, Li23, Li23a}, measuring the incidence rate of mergers in the Universe \citep[e.g.][]{Lotz08-mf, Casteels14, Conselice14, Duncan19, GO23, Nevin23}, as well as in subsamples of galaxies such as AGN \citep[e.g.][]{Villforth14, Chiaberge15, Mahoro19, Hernandez23}, PSBs \citep[e.g.][]{Meus2017, Pawlik18, Saz21, Wilkinson22, Verrico23}, (U)LIRGS \citep[e.g.][]{Sanders96, Murphy96, Ellison13-ulirgs, Pyscho16}, green valley galaxies \citep[e.g.][]{Mahoro19}, and early type galaxies \citep[e.g.][]{Huang22, Giri23}. However, there exists a diversity of results regarding the number of galaxy mergers in different populations sometimes leading to entirely different conclusions. This brings into question the reliability of current merger identification techniques.

Many different methods have been employed for inspecting a galaxy's photometric image and assessing its merger status. The simplest way is visual inspection for telltale disruptions expected from theory to be caused by recent mergers \citep[e.g.][]{Mahoro19,Ellison19,Verrico23}. However, as the amount of photometric data from wide and deep photometric surveys increases, it has become prohibitively time consuming for an individual to visually inspect all images. Furthermore, with human inspection comes inconsistency of opinion \citep{Bickley21, Lambrides21} and therefore a lack of reproducibility. Crowdsourcing the task of numerous visual classifications effort across many individuals \citep[e.g.][]{Lintott08, Lintott11, Willett13, Simmons17} can reduce individual bias, but assumes that the consensus classification is the correct one. Recent advancement in the field of deep learning techniques has made supervised deep learning methods an increasingly common method of automating the classification of galaxy images \citep[e.g.][]{Hocking18, Martin20, Cheng21, Cheng21-CNN-supervised-DES} and specifically distinguishing mergers from non-mergers \citep[e.g.][]{ Bottrell19CNNReal, Bottrell22, Ciprijanovic20, Ferreira20, Ferreira22, Bickley21}. However, deep learning techniques come with a steep learning curve and require a large amount of pre-existing correct labels for training, making deep learning non-trivial to implement.

Another automated approach to identifying mergers that has been implemented for decades is computing non-parametric morphology statistics \citep[e.g.][]{Abraham94, Conselice00, Lotz04, Freeman13, Ferrari15, Wen16, Pawlik16}. Non-parametric morphology statistics are mathematical formulations that take a two-dimensional image and reduce the information therein into a single scalar value. One benefit of using non-parametric morphology statistics as merger indicators is that they are model independent and thus make no prior assumptions as to what the galaxy morphology may be. More important, is that they are relatively quick and easy to understand and implement; at their fundamental level, non-parametric morphology statistics generally consist of very basic mathematical operations (addition/subtraction, multiplication/division). However, subtleties such as where to consider the centre of a galaxy, the extent of the galaxy over which to compute a morphology calculation, and appropriate characterization of the background noise can be particularly onerous. Thankfully, those subtleties have been taken care of for ease of use and consistency across projects by codes such as \texttt{statmorph} \citep{RG19} which have compiled many morphology calculations into a single code package.

Non-parametric morphology statistics have been shown to be sensitive to wavelength \citep{Kelvin12, Vika13, Haussler13,Baes20} and to different image qualities \citep{Lotz04, Lisker08}. Recent works have attempted to characterise the biases introduced by noise properties of the images \citep[e.g.][]{Thorp21} and account for them in several ways \citep[e.g.][]{Deg23, Yu23}, but no consensus has yet been reached. While it doesn't eliminate biases, combining non-parametric morphology statistics together with classical machine learning techniques has been shown to improve merger classification performance. In particular, \citet{Nevin19} uses linear discriminant analysis (LDA) to improve merger identification in Sloan Digital Sky Survey (SDSS) imaging \citep[see also][]{Ferrari15, Ferreira18}. Another method has been to use random forest algorithms with non-parametric morphology statistics as input \citep[e.g.][]{Rose23, GO23}, which was shown to be a more effective classifier than individual non-parametric morphology statistics by \citet{Snyder19}. Despite the biases that may be present, it seems that non-parametric morphology statistics will remain a useful tool for the foreseeable future, employed across wavelength regimes and image qualities. Non-parametric morphology statisitics have been already employed at high redshift using imaging from the James Webb Space Telescope \citep[JWST;][]{Ferreira22-npm, Kartaltepe23, Rose23, Yao23}, and are expected to be used on large scale optical ground based surveys such as Legacy Survey of Space and Time \citep[LSST;][]{Ivezic19, Bignone20}, space based optical surveys such as Chinese Space Station Optical Survey \citep[CSS-OS;][]{Gong19, Yao23}, and even radio imaging of molecular and atomic gas distributions \citep{Deg23, Holwerda23}. 




Since merger features, such as tidal tails or shells, are often diffuse, extended features \citep{Toomre72, Malin83, Quinn84, Johnston02, Wang12, Amorisco15, VC22}, deep imaging is required to assess their presence \citep{Johnston08,Ji14, Duc15, Martin22, DS23}. Several works have shown that observed morphology is not robust in shallow imaging and as a result, fewer mergers are detected \citep{Lotz04, Lisker08}. Another limitation is the seeing of the imaging, as quantified by the full width at half maximum (FWHM) of the point spread function (PSF). Imaging with worse seeing blurs out morphology details important for classification leading to a decrease in classification accuracy \citep{Moore06, Reichard08, Martin22}. Lastly, morphological disturbance has been shown to be most extreme at or near the time of coalescence and slowly fade over time \citep{Lotz08-time, Lotz10-gas, Lotz10-mu, Nevin19, McElroy22}. Thus, fewer mergers are detected at increasing time after coalescence. As a result of these effects, it has been shown that common merger identification techniques are not identifying mergers in their totality \citep{Kampczyk07, Bignone17,  Abruzzo18, Blumenthal20, Lambrides21, McElroy22, Rose23}. Specifically, \citet{Blumenthal20} show that visual identification of mergers recovers only 45\% of interacting galaxies in imaging similar to that of the Sloan Digital Sky Survey \citep[SDSS;][]{York00}. Other works have shown similar levels of completeness using non-parametric morphology statistics; for example, \citet{Rose23} find that Gini-M$_{20}$ recovers 48\% of mergers and \citet{Bignone17} find that 45\% of major mergers satisfy A > 0.35 in SDSS-like imaging. 

While several previous works have quantified the individual effects of depth, resolution, and time since coalescence on observed morphology and merger detection, no one has quantified their combined effect \citep[e.g.][]{Lotz04, Lotz08-time, Moore06, Nevin19, McElroy22}. In this work, we consider six different merger identification methods. Four are known as non-parametric morphology statistics, the other two are examples of classical machine learning techniques that take non-parametric morphology statistics as input. The efficacy of these methods in varying observational conditions can be tested with controlled experiments using data from simulations of galaxy mergers. Unlike real data, simulated data is unaffected by PSF blurring and sky noise and the merger history of galaxies and the properties of the progenitors are known with certainty. Thus, the true image of a galaxy whose merger history is known can be degraded to varying levels of image qualities to test the response of various merger identification metrics.


The paper is structured as follows. In Section \ref{Methods}, we introduce the simulation data and synthetic image generation pipeline. In Section \ref{mergerid}, we define the merger identification methods used in our analysis. In Section \ref{allresults}, we present the merger completeness, false positive rate, and purity across image qualities for each of the merger identification methods and test how the completeness of detected mergers is affected by galaxy properties and orientation of the galaxy with respect to the observer. In Section \ref{discussion}, we discuss the results of our work within the context of the literature and recommend how our results may be useful. Finally, in Section \ref{Summary-TNG}, we summarize the results and conclusions.

\section{Simulation Data}
\label{Methods}
    
    \subsection{Mergers and Controls in IllustrisTNG}
    \label{Illustris}
    
    The IllustrisTNG project is a suite of magneto-hydrodynamical large-box cosmological simulations and is comprised of three different box sizes: (51.7 cMpc)$^3$, (110.7 cMpc)$^3$, and (302.6 cMpc)$^3$ \citep{TNG1,TNG2,TNG5,TNG3,Pillepich19,TNG4}. In this work, we use the highest resolution run of the (110.7 cMpc)$^3$ box simulation, IllustrisTNG100-1 (henceforth TNG100). TNG100 strikes a balance between providing a large sampling of galaxies in realistic cosmological environments (as opposed to idealized merger suites, for example), whilst resolving galaxies with M$_\star > 10^{10}$ $\text{M}_\odot$ with at least $\thicksim 10^4$ particles and an approximate spatial resolution of 0.7 kpc, according to the gravitational softening length of the stellar and dark matter particles \citep{Springel18}. For this reason, several previous works have used TNG100 to study samples of pairs and post-merger galaxies \citep{Patton2020, Hani2020, Quai21, Brown23, BM23}.
    
    One drawback of TNG100 (and indeed the entire IllustrisTNG simulation suite) is that, due to storage limitations, data are saved in coarse time steps called snapshots which vary in temporal size according to the redshift of the simulation at the time. The average time between snapshots at $z<0.2$ of the simulation is 169 Myr, which limits our ability to track galaxy properties on timescales smaller than this, but still grants the power of determining a temporally coarse, but complete, assembly/merger history of all galaxies in the simulation.
    
    The first step in generating the assembly/merger histories of galaxies in TNG100 is to identify individual dark matter halos at each snapshot using a friends-of-friends algorithm \citep{Davis85} on the dark matter particles, requiring a minimum of 32 particles to constitute a halo. After dark matter halos have been established, gas and star particles are assigned to the halo to which the nearest dark matter particle belongs. Substructure within the halo (i.e. subhalos) are identified using an extension of the \texttt{SUBFIND} algorithm \citep{Springel01, Dolag09} and temporally connected through snapshots of the simulation using the \texttt{SUBLINK} algorithm. These algorithms have been applied to TNG100 following the methods of \citet{RG15} and the out falling data is publicly available in the IllustrisTNG catalogues\footnote{https://www.tng-project.org/data/docs/specifications/}.

    A natural consequence of hierarchical growth of galaxies and tracing galaxies through cosmological time with \texttt{SUBLINK} is the generation of merger trees. Merger trees identify when two subhalos become one and allow you to trace back the progenitor galaxies to understand their individual properties. The publicly available merger trees of TNG100 were parsed by \citet{BM23b} providing a catalogue containing the time since the most recent merger occurred and the mass ratio of that merger, among other properties. From this catalogue of merger properties in TNG100, we identify 426 low-redshift isolated post-mergers that meet the following requirements:

\begin{itemize}
    \item \emph{Recent:} the snapshot at which the post-merger is synthetically observed is the first snapshot of the simulation in which the two progenitors have coalesced. Due to the temporal resolution of the simulation, this corresponds to a post-merger timescale of $0 < t_\text{post-merger}\lesssim$ 200 Myr. The temporal resolution of TNG100 is too coarse for a detailed study of how merger features fade over time. For now, we only consider the most recent mergers in the simulation to assess the \emph{best-case-scenario} for merger detection.
    \smallskip

    \item \emph{Low-redshift:} the mergers must occur at $z\lesssim0.2$ of the simulation. At higher redshifts, there is significant evolution in galaxy properties that affect their morphologies causing them to statistically differ from the $z\sim0.1$ galaxies of the real Universe.
    \smallskip
    
    \item \emph{Massive:} The post-merger remnants must have a total stellar mass greater than $10^{10}$ M$_\odot$. This mass ensures the merger remnant is not affected by spurious heating from dark matter particles \citep{Ludlow21,Ludlow23} and that both progenitors are well resolved (in terms of number of particles), as long as a minimum mass ratio between the progenitors is required. 
    \smallskip
    
    \item \emph{Significant mass ratio:} the two progenitors of the post-mergers must have a mass-ratio less disparate than 1:10 (i.e. $\mu > 0.1$). This ensures the smallest progenitor would have a total stellar mass of $10^9 \text{M}_\odot$ corresponding to $\thicksim10^3$ star particles.
    \smallskip
    
    \item \emph{Isolated:} the post-merger galaxy must not merge with another galaxy one-tenth its stellar mass or greater in the subsequent snapshot and have no massive ($M_{\star} > 10^9 M_\odot$) neighbouring galaxy within 100 kpc. This ensures the post-mergers are isolated and not being influenced by an additional pair-phase interaction.
    \smallskip
    
\end{itemize}


    \begin{figure}
    \centering
        \includegraphics[width=1\linewidth]{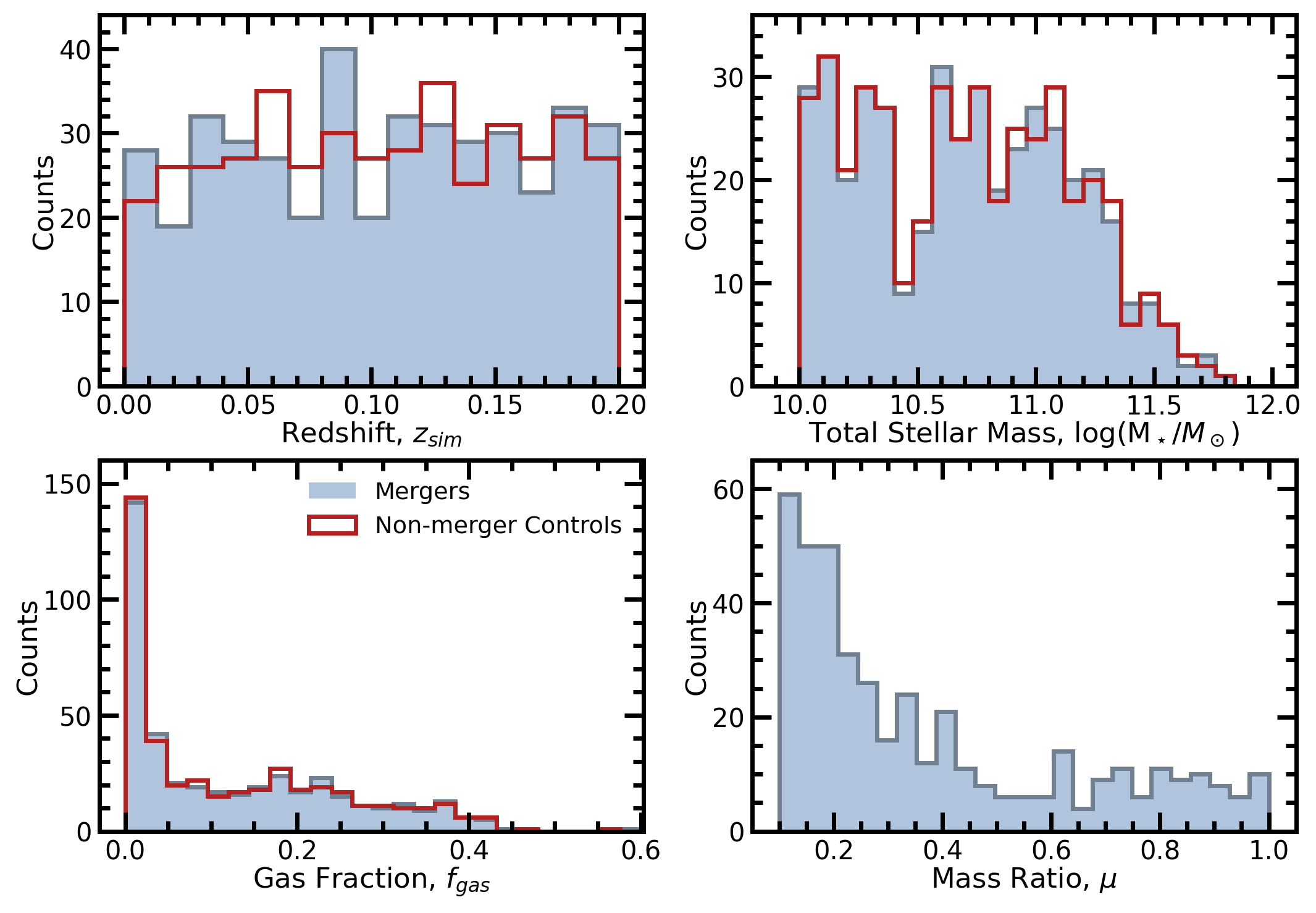}
        \caption{The distributions of the 424 TNG100 post-mergers in simulation snapshot, total stellar mass, gas fraction, and mass ratio (light blue filled histograms) along with the non-merger control sample which was matched in snapshot, total stellar mass, and gas fraction (red open histograms).}
        \label{MergerDists}
    \end{figure}

    For a comprehensive assessment of the efficacy of various merger identification methods, it is integral to quantify how frequently they misclassify non-mergers. Testing this requires a complementary non-merger control sample. The pool from which non-merger controls are selected include all of galaxies in TNG100 at $z_\text{sim} < 0.2$ that have no nearby companion galaxies within 100 kpc, have not undergone a merger with another galaxy 1\% of its mass or greater (i.e. $\mu > 0.01$) within 2 Gyr, and will not merge within the subsequent snapshot. There are 57,654 galaxies that meet these criteria.

    Each of the 426 post-mergers in TNG100 are matched to one non-merger from this control pool in simulation redshift, total stellar mass, and gas fraction within two half mass radii\footnote{Defined as $f_\text{gas} = \frac{M_\text{gas}(r < 2R_{1/2})}{M_\star(r < 2R_{1/2}) + M_\text{gas}(r < 2R_{1/2})}$, the gas fraction measured by taking all stellar and gas mass within two half mass radii ($R_{1/2}$) incorporates only the gas that can reasonably impact star formation and thus morphology of the galaxy on short timescales and excludes the reservoir of hot gas that may surround the galaxy.}. Each matched control is required to be within $\pm 1$ snapshot (i.e. a small window of redshift), with a matching tolerance of $\Delta \text{log}(M_\star) <0.1$ dex and $\Delta f_\text{gas} <0.05$. The best-matched control is then taken to be that which minimizes the distance in log($M_\star$)-$f_\text{gas}$ space to the merger and are selected iteratively without replacement from the pool of possible controls. Specifically, each post-merger with a stellar mass, $M_{\star\text{, PM}}$, and a gas fraction, $f_{\text{gas, PM}}$, receives one matched control which has a stellar mass, $M_{\star\text{, C}}$, and gas fraction, $f_{\text{gas, C}}$, that minimizes the following equation:

    \begin{equation}
        \sqrt{\left[\text{log}\left(\frac{M_{\star\text{, PM}}}{M_\odot}\right) - \text{log}\left(\frac{M_{\star\text{, C}}}{M_\odot}\right)\right]^2 + \left[f_{\text{gas, PM}} - f_{\text{gas, C}}\right]^2}
    \end{equation}

    424 of the 426 mergers are successfully matched to controls within these tolerances, typically well within the allotted tolerances; on average the mergers and controls differ by $\Delta \text{log}(\frac{M_\star}{M_\odot}) = 0.0038$ dex and $\Delta f_\text{gas} = 0.0035$. The two mergers without matched controls are removed from the sample, leaving a final sample of 424 mergers and 424 non-merger controls. The simulation redshift, stellar mass, and gas fraction distributions of the merger sample (blue) and non-merger controls (red) are shown in Figure \ref{MergerDists}.

    \subsection{Synthetic Imaging}
    \label{SyntheticImages}

    \subsubsection{Radiative Transfer with SKIRT9}

    Our ability to translate the conclusions drawn from the analysis presented in this work to the real Universe is ultimately dependent upon the simulation's ability to reproduce and represent realistic morphologies. While the IllustrisTNG simulations are tuned to approximately reproduce certain characteristics of the observable universe such as the galaxy mass function and sizes at z = 0, the simulations have not been tuned to reproduce observed galaxy morphologies. However, recent works have shown that the physics prescribed by the simulation give rise to realistic galaxy morphologies, particularly when radiative transfer simulations are used to incorporate the effects of dust attenuation \citep{RG19, Tacchella19}. Therefore, in this work, we use the data from the TNG100 simulation as input to SKIRT \citep{Baes11, Camps15, Baes20}, a publicly available radiative transfer simulation code, to create realistic images from the simulation. Our SKIRT pipeline is essentially the same as that described in \citet{Bottrell23}, with the only major difference being that all mergers are simulated at a fixed redshift of 0.1 rather than at the simulation redshift. Readers looking for a detailed description of the TNG100-to-SKIRT pipeline are encouraged to read \citet{Bottrell23}, but the most important points for this work are described here.

    The inputs to SKIRT from the simulation are the location and properties of the gas and star particles belonging to the halo to which the target galaxy (TNG100 subhalo) belongs. The star particle data are used to emit photon packets into the SKIRT simulation with wavelengths and luminosities according to a prescribed stellar spectral model. For star particles with ages greater than 10 Myr, we use the model spectra library from \citet{BC03} and for star particles with ages less than 10 Myr, we use the \texttt{MAPPINGSIII} library \citep{Groves08}, which better accounts for stellar luminosities in young stellar envelopes. Since dust density is not tracked explicitly in TNG100, it is determined following the method of \citet{Popping22} whereby we use a redshift-independent dust model in which the dust-to-metal mass ratio in the gas scales with its metallicity. The model is motivated by empirical scalings between these properties derived for local galaxies by \citet{RR14}. The distribution of dust grain sizes is set according to a Weingartner and Draine dust mix \citep{WD01} with properties tuned to a Milky Way extinction curve. The photon packets then travel outwards through the galaxy, being possibly scattered or attenuated by the dust, and recorded by the camera instrument outside the galaxy. 

    Upon arrival of the stellar light to the instrument, the light is redshifted and dimmed as if the galaxy was at a redshift $z = 0.1$ from the observing instrument. We choose to make broadband photometric images with this light (rather than spectral datacubes) in the $r$-band, to trace the optical light from stars in the galaxy. Specifically, we use the CFHT MegaCam $r$-band filter response\footnote{https://www.cadc-ccda.hia-iha.nrc-cnrc.gc.ca/en/megapipe/docs/filt.html} as a representative mid-optical passband. The photometric images are generated with a field of view equal to either two half mass radii of the dark matter halo or twenty half mass radii of the stellar mass distribution, whichever is larger, and with a pixel scale of 0.1 kpc per pixel.

    For each galaxy, we generate four unique images using cameras situated at the vertices of a tetrahedron oriented with respect to the simulation box and not any specific property of the galaxy (i.e. four random orientations, as if observed from Earth). This brings our total number of SKIRT images to 3,392 for the mergers and controls combined.

    \subsubsection{Adding Noise and Atmospheric Blurring}

    The generated SKIRT images have a very high resolution pixel size of 0.1 kpc / pix ($\thicksim 0.05$ arcsec / pix at $z = 0.1$), no sky noise, and no atmospheric blurring. To make the images more realistic to observations, we first down-sample the images to a fixed pixel scale of 0.1 arcsec / pix following the methods of the \texttt{RealSim}\footnote{github.com/cbottrell/RealSim} code \citep{Bottrell19CNNReal}. 

    To emulate the effect of atmospheric blurring, the re-binned SKIRT image is convolved with a PSF. PSFs, in general, account for not only all aberrations of the light including those caused by the atmosphere but also the instrumentation itself. The intent of this work is to generalize beyond any specific instrument. Therefore, we use a simple 2-D Gaussian profile to model the blurring as characterized by its FWHM.

    After applying a PSF, the image is then co-added with an equally sized field of Gaussian random sky noise. The sky flux in each pixel is drawn from a Gaussian profile with a mean of 0 (i.e. the image is background subtracted) and a standard deviation according to a pre-determined depth (e.g. $\sigma_\text{sky}$ = 26 mag arcsec$^{-2}$; see Table \ref{DepthConv} and Section \ref{interpretation} for a discussion of how this relates to other definitions of depth). Of course, Gaussian sky noise does not encompass all the complexities of real skies, particularly those in surveys which may have correlated noise due to the survey observation patterns and the data reduction pipeline. For the experiments presented in this work, real skies can't be drawn from surveys as is done in other works \citep[e.g.][]{Bottrell19CNNReal,Bickley21,Ferreira22} since the experiments require full and individual control over both the depth and the resolution of the synthetic images. Nonetheless, the image construction serves as an example set from which the trends in PSF and sky noise on merger classification can be investigated.

\begin{figure*}
        \centering
        \includegraphics[width=1\linewidth]{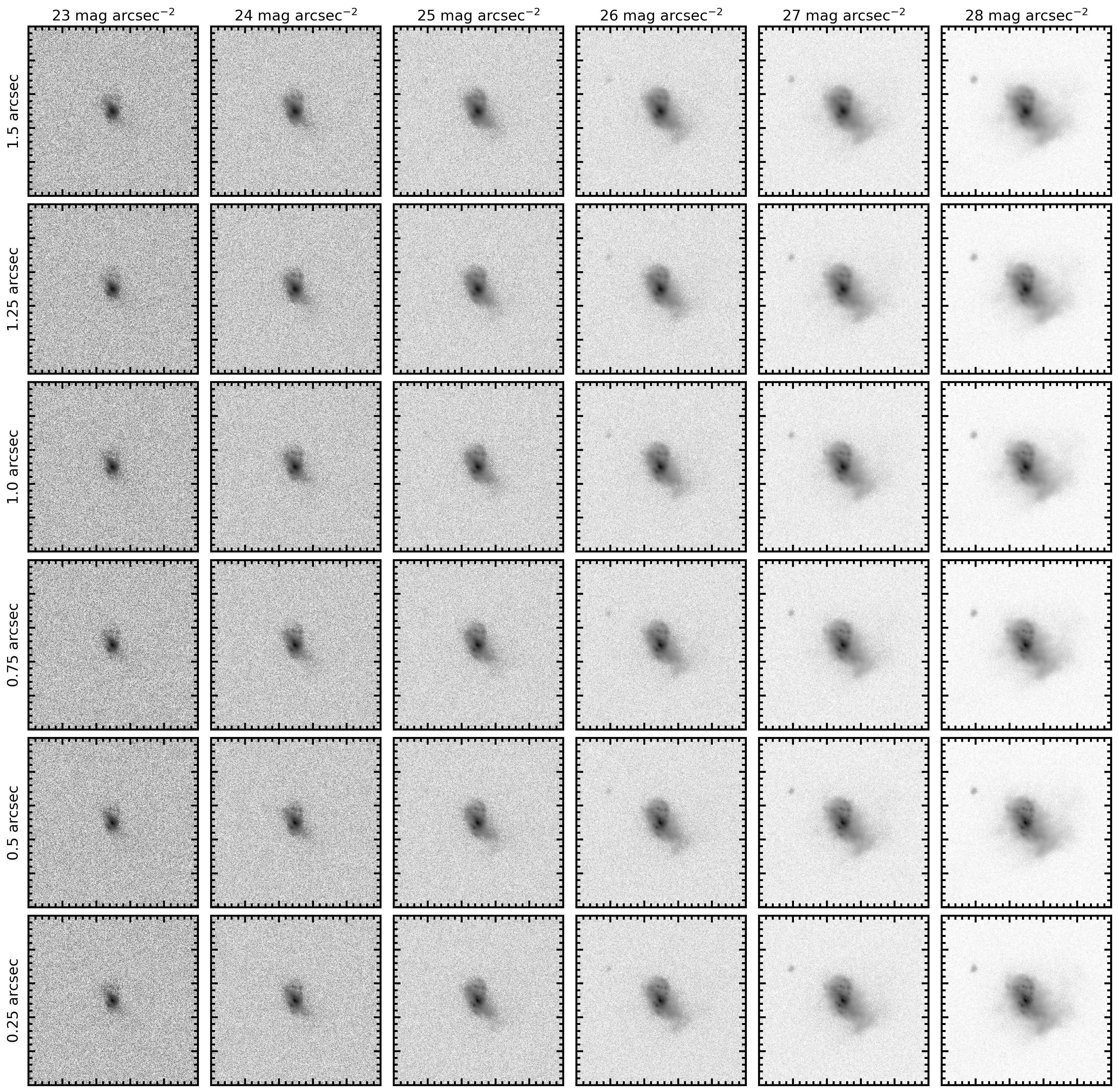}
        \caption{Thirty-six degraded images of one idealized synthetic image at one viewing angle, each displayed on an equivalent logarithmic scaling. From left to right, the images improve in depth, ranging from 23-28 mag arcsec$^{-2}$. From top to bottom, the images have reduced atmospheric blurring, ranging from 1.5-0.25 arcsec. Therefore, the worst image quality used in this experiment is in the top left panel and the best is in the bottom right panel. The galaxy is a post-merger drawn from TNG100 (snapshot 88, subhalo 465168) that has a total stellar mass of 3.5$\times10^{10}M_\odot$, a gas fraction of 0.179, and has recently undergone a merger with a mass ratio of 0.177}
        \label{RealExample}
    \end{figure*}

    For the experiment at hand, the SKIRT image of each galaxy (and at each viewing angle) is degraded to a 6$\times$6 grid of varying image qualities spanning six depths from 23-28 mag arcsec$^{-2}$ and six PSFs with FWHMs ranging from 0.25-1.5 arcsec. The ranges in seeing and depth are selected to span from slightly worse than the SDSS \citep{York00} to slightly better than that of the expected 10-year depth of the LSST \citep{Laine18, Ivezic19, Brough20, Martin22}. We do not reach the resolution of space-based diffraction-limited telescopes such as Hubble since the TNG100 simulation is too low resolution. However, in addition to the 6$\times$6 grid of image qualities, we generate an ``ideal" image for each galaxy which has the same pixel scale but no atmospheric blurring effects and 30 mag arcsec$^{-2}$ sky noise. In total there are $(424+424) \text{ galaxies} \times 4 \text{ viewing angles} \times (36+1) \text{ image qualities} = 125,504$ unique galaxy images used throughout this work. In Figure \ref{RealExample}, we present all thirty-six combinations of PSF and depth for one example post-merger drawn from TNG100 (snapshot 88, subhalo 465168). This post-merger has a total stellar mass of 3.5$\times10^{10}M_\odot$, a gas fraction of 0.179, and has recently undergone a merger with a mass ratio of 0.177.

\section{Merger Identification Methods}
\label{mergerid}
    
    In this work, we consider six different merger identification methods. Three are stand-alone non-parametric morphology statistics, the fourth is a linear combination of two non-parametric morphology statistics and the latter two are classical machine learning methods which use non-parametric morphology statistics as input. The non-parametric morphology statistics themselves are computed using the Python package \texttt{statmorph} \citep{RG19}. In this section, we describe both the individual non-parametric morphology statistics and the two supervised machine learning methods used in this work, all of which have been implemented by previous works to identify mergers.
    
    \subsection{Non-parametric Morphology Statistics}
    \label{NPMS}
    
    \subsubsection{Asymmetry}
    
    Asymmetry \citep{Abraham96, Conselice00}, $A$, quantifies the azimuthal asymmetry of a galaxy's light profile. To extract this information, the absolute difference between an image and its 180$^\circ$-rotated counterpart is summed on a pixel-by-pixel basis and normalized by the total absolute flux of the original image. To account for contributions to the asymmetry from the background noise, the average asymmetry of the background is subtracted from the total:
    
    \begin{equation}
        A = \frac{\Sigma |I_0 - I_{180}|}{\Sigma |I_0|} - A_{\text{bgr}},
        \label{asym}
    \end{equation}
    
    \noindent where $I_0$ is the flux of a pixel in the original image, $I_{180}$ is the flux of the same pixel after the image has been rotated by 180$^\circ$ about the centre of the galaxy, determined by minimizing the value of $A$, and $A_\text{bgr}$ is the average asymmetry of the background. The sum is carried out over all pixels within a circular aperture with radius of 1.5 Petrosian radii.

    Low values of asymmetry indicate the galaxy is very azimuthally symmetric, a common feature of early-type galaxies with spheroidal morphologies \citep{Conselice03}. Spiral galaxies inherently have slightly elevated asymmetries due to naturally occurring asymmetric features like dust lanes and clumpy star formation \citep{Conselice03}. Higher asymmetry values are common amongst galaxies exhibiting strong merger and post-merger signatures; \citet{Conselice03} suggest galaxy mergers are those with $A > 0.35$. 

    
    \subsubsection{Outer Asymmetry}
    \label{Ao}

    Outer asymmetry \citep{Wen16}, $A_O$, is defined in the same way as asymmetry (see Equation \ref{asym}) but the sum is computed over an annulus aperture rather than a circular aperture with a radius of 1.5 Petrosian radii. The annulus over which the sum is computed has an inner boundary equal to an ellipse within which 50\% of the total light of the galaxy is contained and an outer boundary equal to the maximum radius of a pixel belonging to the galaxy. Since the central region of a galaxy is often bright and symmetric, excluding the central region from the asymmetry sum allows for $A_O$ to be more sensitive to faint and asymmetric tidal features. \citet{Wen16} note that most galaxies have $A_O < 0.6$, suggesting that galaxies above this threshold are mostly merging galaxies. We therefore take the default merger threshold to be $A_O > 0.6$.
    
    \subsubsection{Shape Asymmetry}
    \label{As_methods}
    
    Shape asymmetry \citep{Pawlik16}, $A_S$, is also defined in the same way as asymmetry (see Equation \ref{asym}) but instead of each pixel having an intensity, $I$, each pixel is given a binary value: 1 if the pixel belongs to the galaxy, 0 otherwise. The binary mask used to measure shape asymmetry is distinct from the binary segmentation map provided to \texttt{statmorph} as input and is generated internally by \texttt{statmorph} following the method described in \citet{Pawlik16}. The point around which the image is rotated is the same as the point in the asymmetry measurement which minimizes the light-weighted asymmetry. 
    
    By calculating the asymmetry of the binary mask rather than the flux of the image itself, equal weight is given to pixels belonging to both the faint and bright regions of the galaxy. Comparatively, standard light-weighted asymmetry weights the central region of a given galaxy proportionally to its flux relative to the faint tidal features around it, which often differ by several orders of magnitude. Hence, the shape asymmetry statistic is more sensitive to low surface brightness features, such as faint tidal streams. Following \citet{Pawlik16} and \citet{Wilkinson22}, we take the default merger threshold to be $A_S > 0.4$.
    
    \subsubsection{Gini-M$_{20}$ Merger Statistic}
    \label{GM20}
    
    The Gini coefficent \citep{Lotz04}, G, is defined as the mean of the absolute difference of the light curve from a uniform distribution where the variable $X$ describes the flux in each pixel and is ordered from lowest to highest flux:
    
    \begin{equation}
        \text{G} = \frac{1}{\overline{X}n(n-1)}\sum_i^n(2i-n-1)|X_i|,
        \label{G}
    \end{equation}
    
    \noindent where $n$ is the number of pixels associated with the galaxy and $\overline{X}$ is the mean of flux of the pixels belonging to the galaxy.
    
    The Gini coefficient is independent of the location of the brightest pixel and tends towards unity if the light from the galaxy is concentrated in a small number of pixels. If all of the galaxy's light were to come from a single pixel, G $= 1$, and if the light is evenly distributed across every pixel in the galaxy, G $= 0$. For a galaxy with a recent burst of star formation in the central regions, one might expect an increase in the Gini coefficient. 
    
    Before defining M$_{20}$, we first introduce the second moment of the total light distribution, M$_\text{tot}$. The second moment of the light distribution, related to the spatial variance of the light distribution, is the summation of the flux of each pixel $f_i$ multiplied by the squared distance from the centre:
    
    \begin{equation}
        \hspace{1mm}{\text{M}_\text{tot}} = \sum_i^n\text{M}_i = \sum_i^nf_i\left[ (x_i - x_c)^2 + (y_i - y_c)^2\right],
        \label{Mtot}
    \end{equation}
    
    \noindent where $x_c$ and $y_c$ are x- and y-coordinates of the centre, determined by selecting the x- and y-coordinates that minimize M$_\text{tot}$.
    
    M$_{20}$, then, is defined as the second moment of the brightest pixels that produce 20\% of the galaxy's light, normalized by the total second moment of the galaxy. If pixels are ordered from highest to lowest flux, M$_{20}$ is calculated as:
    
    \begin{equation}
        \text{M}_{20} = \text{log}_{10} \left( \frac{\sum_{i}\text{M}_i}{\text{M}_\text{tot}} \right), \text{while} \sum_if_i < 0.2f_\text{tot}.
        \label{M20}
    \end{equation}
    
    In tandem, the Gini coefficient and M$_{20}$ can be used to identify galaxies with large portions of the total light profile contained within a small number of spatially separated pixels. Following \citet{Snyder15-mergerstat} and \citet{RG19}, we compute the Gini-M$_{20}$ merger statistic as:

    \begin{equation}
        S(G,M_{20}) = 0.139\text{M}_{20} + 0.990G - 0.327
    \end{equation}
    
    By this equation, mergers are defined by having S(G, M$_{20}$) $> 0$ and thus we take this to be the default threshold in this work.

    \subsection{Supervised Machine Learning Methods}

    In addition to the individual non-parametric morphologies described in Section \ref{NPMS}, we implement two types of supervised classical machine learning methods: linear discriminant analysis and random forest. These supervised machine learning methods can combine the non-parametric morphology statistics and use the information together holistically to infer a merger or non-merger classification.

    By nature of being supervised methods, both machine learning techniques require a training and test set of inputs and targets. The training set is the sample of galaxies that will be used to train the model. The test set is a separate sample of galaxies that the model never sees during training and can therefore be used to assess how the model performs on completely new cases. We randomly select 70\% of the unique galaxy identifiers from the 848 mergers and controls and assign them to the training set. The remaining 30\% are the test set. The split is conducted using the \texttt{train\_test\_split} function from the \textsc{scikit-learn} Python module \citep{Pedregosa11}. The 70/30 split maximizes the number of galaxies that the model has to learn from and generalize over, while reserving a large enough number for robust statistics during assessment of the model. By randomly sorting galaxies into training and test sets, then assigning the morphology data from all four viewing angles to training and test sets according to the initial galaxy sorting, we avoid cases where the same galaxy ends up in the training and test set, but viewed from different orientations. The input to the models will be values of the individual non-parametric morphology statistics $A$, $A_O$, $A_S$, G, and M$_{20}$. The target is a binary value indicating merger or non-merger control.

    \subsubsection{Linear Discriminant Analysis Classifier}

    LDA is a method which takes $n$-dimensional continuous input data and seeks to find an $m-1$ number of hyperplanes that separate a cloud of input vectors into $m$ classes. In our case, the input dimensionality is five ($n = 5$), consisting of $A$, $A_O$, $A_S$, G, and M$_{20}$, all of which are continuous variables. Therefore, each galaxy in the training set becomes a point in 5-dimensional space, possessing a pre-existing label of merger or non-merger. Since we are only seeking a classification between two classes (merger and non-merger), $m = 2$, and the LDA will find one optimized hyperplane which separates the merger and non-merger classes. Once the optimal hyperplane has been determined using the training data, the positions of the test data in 5-dimensional space with respect to the hyperplane give rise to merger/non-merger predictions. For a more detailed description of LDA and its application to classifying galaxy mergers, we refer readers to \citet{Nevin19}.

    In this work, we use the default \texttt{LinearDiscriminantAnalysis} model from \textsc{scikit-learn}. We tried optimizing the hyperparameters with \texttt{GridSearchCV} to see if we could improve performance. However, the change in performance was negligible and so we kept to the default for simplicity.

    Our final output is the probability of classification by the LDA model, $P_\text{LDA}$(merger), garnered using the \texttt{predictproba\_} function. The natural merger threshold is the threshold at which merger prediction is more likely than the alternative. In other words, we take our default merger threshold to be $P_\text{LDA}(\text{merger}) > 0.5$.

    \subsubsection{Random Forest Classifier}

    A random forest classifier is a model in which classification arises from an ensemble of decision trees. A single decision tree poses a series of questions, restricted to the domain of the input data, with the goal of classifying the target as one class or the other, in our case merger or non-merger. The model optimizes to ask the most potent questions of the input morphology data for merger identification. Since a difference in the early part of a decision tree can produce compounding differences along the branches of the tree, noise in the model and a lack of generalization is combated by developing a number of decision trees. Each decision tree is given a slightly different set of galaxies from the training set in a bootstrap fashion. Finally, the output from this collection of trees, or forest, averaged together to get a probability of a merger. Several previous works have used non-parametric morphology inputs to random forests, for the purpose of classifying mergers and non-mergers \citep[e.g.][]{Snyder19, GO23, Rose23}.

    Specifically, we use the \texttt{RandomForestClassifier} model from \textsc{scikit-learn}, with hyperparameters tuned using \texttt{GridSearchCV} for the highest quality imaging. The only changes to the hyperparameters from the default \texttt{RandomForestClassifier} model are $\texttt{n\_estimators} = 1000$, $\texttt{criterion} = \texttt{'entropy'}$, $\texttt{min\_samples\_leaf} = 10$, and $\texttt{min\_samples\_split} = 25$.

    Our final output is the probability of classification by the random forest, $P_\text{RF}$(merger), garnered using the \texttt{predictproba\_} function. The natural merger threshold is the threshold at which merger prediction is more likely than the alternative. In other words, we take our default merger threshold to be $P_\text{RF}(\text{merger}) > 0.5$.

\section{Results}
\label{allresults}

    In this section, we seek to understand and quantify the factors affecting one's ability to identify mergers in a given sample. This will be tested by measuring the non-parametric morphology statistics useful for identifying mergers for the sample of recent post-merger galaxies drawn from IllustrisTNG that intrinsically have a realistic distribution of galaxy merger properties such as orbital parameters and mass ratio of the progenitors. The ability of a given merger identification method to identify mergers will be assessed using the completeness of the recovered mergers (equivalently known as true positive rate, recovery fraction, recall, or sensitivity), the false positive rate, and purity (equivalently known as positive predictive value). Completeness is measured as the number of true mergers identified correctly by the merger identification method ($N_\text{correct mergers}$) divided by the number of galaxies in the total merger sample($N_\text{total mergers}$):

    \begin{equation}
        \text{Completeness} = \frac{N_\text{correct mergers}}{N_\text{total mergers}}.
        \label{CompletenessEq}
    \end{equation}

    Likewise, the false positive rate (FPR) is measured as the number of non-mergers incorrectly classified as mergers ($N_\text{incorrect non-mergers}$) divided by the number of galaxies in the total non-merger sample ($N_\text{total non-mergers}$):

    \begin{equation}
        \text{False Positive Rate (FPR)} = \frac{N_\text{incorrect non-mergers}}{N_\text{total non-mergers}}.
        \label{FPREq}
    \end{equation}
    
    Lastly, purity is defined as the ratio of the number of correctly identified mergers to the total number of detected ``mergers" (i.e. sum of true and false positives): 
    
    \begin{equation}
        \text{Purity} = \frac{N_\text{correct mergers}}{N_\text{correct mergers} + N_\text{incorrect non-mergers}}.
        \label{PurityEq}
    \end{equation}
    
    Unlike completeness and false positive rate, purity depends on the prevalence of true positives in a given sample. This is important to merger identification because mergers are rare, particularly in the low-redshift universe \citep{Lotz11}. Merger rates also evolve with redshift \citep{Duncan19, Ferreira20, Whitney21} and vary between sub-classes of galaxies, which would change the purity of mergers identified by the methods tested here, but not the completeness or false positive rate. Thus, the purity presented in this work -- derived using a balanced dataset of equal numbers of mergers and non-mergers -- should not be assumed to be applicable to any given galaxy sample. Purity is included in the results to help give the reader a sense of how well a metric can distinguish between mergers and non-mergers.

    For each quality of imaging shown in Figure \ref{RealExample}, we measure the non-parametric morphology statistics and merger probabilities from the LDA and random forest methods for the degraded synthetic galaxy images. In this section, we first present and discuss the completeness, false positive rate, and purity results for a single image quality, namely the idealized imaging with a depth of 30 mag arcsec$^{-2}$ and no PSF blurring in Section \ref{Results-Ideal}. Then, we discuss how those results are affected by the depth and PSF blurring of the images, respectively and in tandem in Section \ref{Results-IQ}. We then go on to demonstrate how several galaxy properties and the orientation of a galaxy with respect to the observer can affect merger observability in Sections \ref{Results-props} and \ref{disc-ViewAngExp}.

\subsection{Merger Identification in Ideal Imaging}
\label{Results-Ideal}

\begin{figure*}
        \centering
        \includegraphics[width=1\linewidth]{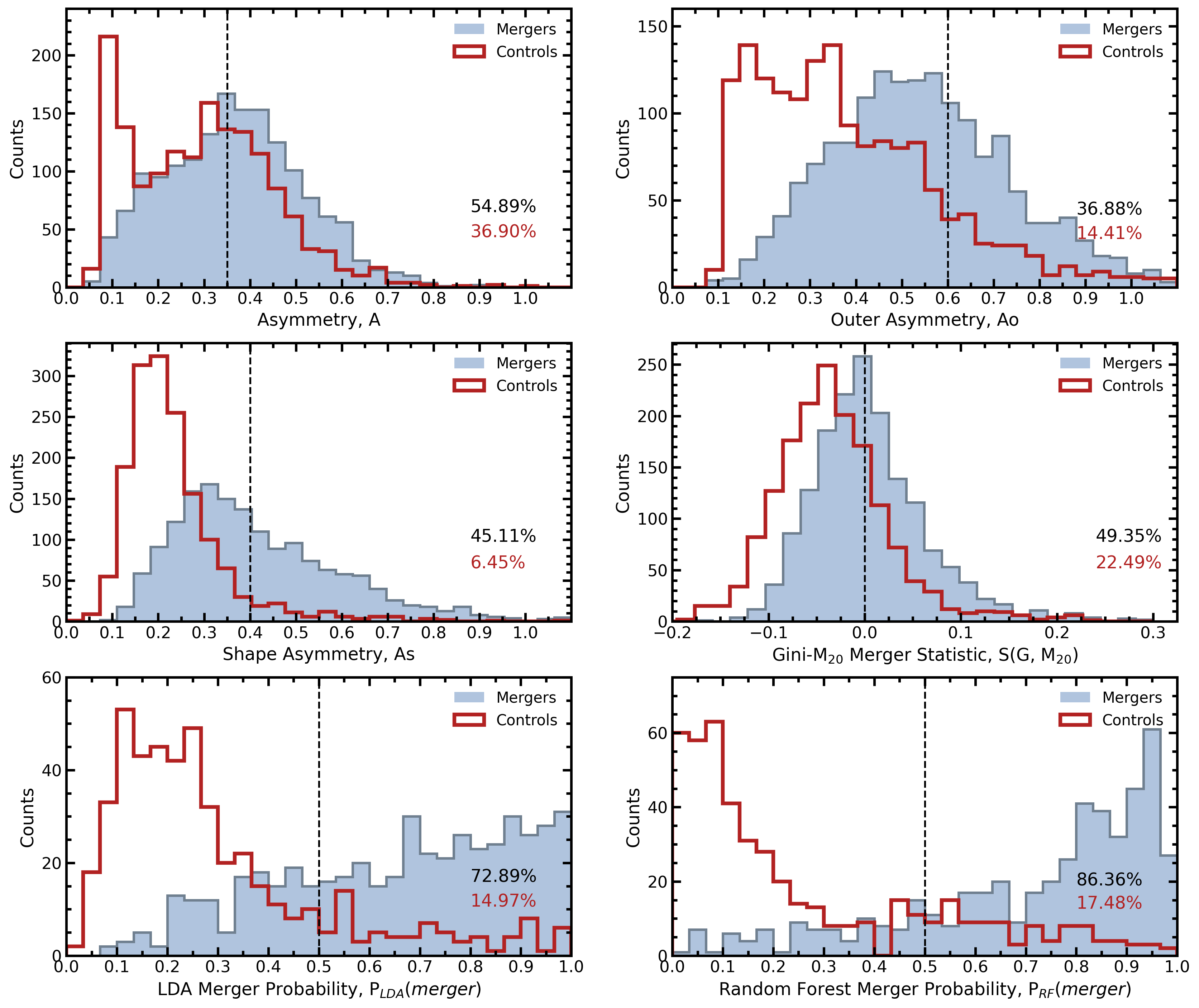}
        \caption{The distribution of four non-parametric morphology statistics commonly used to identify mergers and the probability from the LDA and random forest methods as applied to a sample of IllustrisTNG mergers in ``ideal" imaging (no atmospheric blurring and a sky noise of 30 mag arcsec$^{-2}$). The blue histograms are the post-merger sample and the red open histograms are the non-merger control sample. Vertical dashed lines demarcate the default merger detection threshold for each statistic, above which a galaxy is considered a merger. The percentage of post-mergers above each threshold (i.e. the completeness) is written in black text on each panel. The percentage of non-mergers above each merger threshold (i.e. the false positive rate) is written in red text on each panel. In the case of the LDA and random forest, only post-merger probabilities of the post-mergers and controls from the test set are shown.}
        \label{OneIQ-dists}
    \end{figure*}

    For each combination of PSF blurring and depth (see Figure \ref{RealExample}) and the ideal imaging, the 3,392 synthetic images of galaxy mergers and non-mergers at that image quality are processed with \texttt{statmorph} which provides non-parametric morphology statistics for each of the synthetic images. In Figure \ref{OneIQ-dists}, we present the distributions of four non-parametric morphology statistics and probabilities from the machine learning methods for the 3,392 ideal images of mergers and non-mergers from TNG100. The vertical lines in each panel of Figure \ref{OneIQ-dists} demarcate the default threshold for each metric above which galaxies are considered detected mergers. 

    Considering first the non-parametric morphology distributions of the merger sample presented in blue histograms in the top four panels of Figure \ref{OneIQ-dists}, our results demonstrate that even in the ideal quality images, more often than not mergers have morphologies below the merger thresholds of the non-parametric morphology statistics. This is quantified by the percentage of mergers which are above the default thresholds (i.e. completeness of merger sample) reported in black in the bottom right of each panel. Specifically, asymmetry has the highest completeness with 54.9\%; outer asymmetry has the lowest completeness with 36.9\%; shape asymmetry achieves a completeness of 45.1\%; and the Gini-M$_{20}$ merger statistic has a completeness of 49.4\% of mergers.

    The red open histograms in the top four panels of Figure \ref{OneIQ-dists} demonstrate that in ideal imaging a significant fraction of the non-merger control galaxies have non-parametric morphologies above the merger thresholds. This is quantified by the percentage of non-merger controls which are incorrectly classified as mergers (i.e. the false positive rate) reported as a percentage in red in the bottom right of each panel. Folding in completeness and the false positive rate, we also compute the purity. While asymmetry has the highest completeness (54.9\%), it also has the highest false positive rate (36.9\%), leading to a purity of 60.3\%. Unlike the merger sample, the asymmetry distribution of the non-merger controls is distinctly bimodal; the peak near zero is driven by controls with low star formation rates and the higher, broader peak is driven by controls with higher star formation rates. This is true for observations of real galaxies to some extent, but the seeing reduces the asymmetry contribution from bursty star-formation. In these idealized images there is no such blurring, leading to higher asymmetry measurements and thus increased completeness and false positive rates for the mergers and controls. This bimodal asymmetry distribution is not seen in the lower quality images. Outer asymmetry has a lower false positive rate than standard asymmetry (14.4\%), giving a purity of 72.3\%. Shape asymmetry, which is wholly agnostic to any possible unnaturally bursty star formation has the lowest false positive rate (6.5\%) and the highest purity (87.7\%) of any of the individual non-parametric morphology statistics. Finally, the Gini-M$_{20}$ merger statistic has a false positive rate of 22.5\%, giving a purity of 69.1\%.
    
    Despite the relatively poor performance of the individual non-parametric morphology statistics, when $A$, $A_O$, $A_S$, G and M$_{20}$ are combined together using a random forest or linear discriminant analysis method, both the completeness and purity of the identified mergers increases greatly. Focusing now on the bottom two panels of Figure \ref{OneIQ-dists}, the LDA and random forest post-merger probability distributions of the mergers and non-mergers are much more distinctly separated than any of the individual non-parametric statistics. Indeed the LDA has a completeness of 72.9\%, false positive rate of 15.0\% and a purity of 83.2\%. The random forest performs even better, with a completeness of 86.4\%, a false positive rate of 17.5\% and a purity of 83.4\%. 

    The merger identification results based on ideal imaging presented in this subsection demonstrate that combining non-parametric morphology statistics together using a classical machine learning techniques can nearly double the completeness of individual statistics, whilst reducing the false positive rate. In the following subsection, we will show that this result is qualitatively ubiquitous across all image qualities tested. Thus, we advise readers to exercise caution when using a non-parametric morphology statistic unilaterally for merger identification. Instead, we recommend training a LDA or random forest algorithm based on labelled training data specific to your dataset. This can be done with very simple code and relatively few pre-existing labels (see Appendix \ref{Nlabels}).


\subsection{The Effect of Depth and Seeing}
\label{CompResults}
\label{Results-IQ}

Since synthetic images of the entire merger sample are generated for each image quality and processed with \texttt{statmorph}, the non-parametric morphology distributions (as in Figure \ref{OneIQ-dists}) of the entire sample exists for all 36 image qualities. Rather than showing the distributions of the morphology statistics at every image quality, we extract the completeness, false positive rate, and purity information from the distributions, so that trends with image quality can be seen in a more compact form. What follows is a detailed description of the observed trends between completeness, false positive rate, and purity with changing quality of imaging.

\subsubsection{Asymmetry}
\label{Aresults}

\begin{figure*}
    \centering
    \includegraphics[width=1\linewidth]{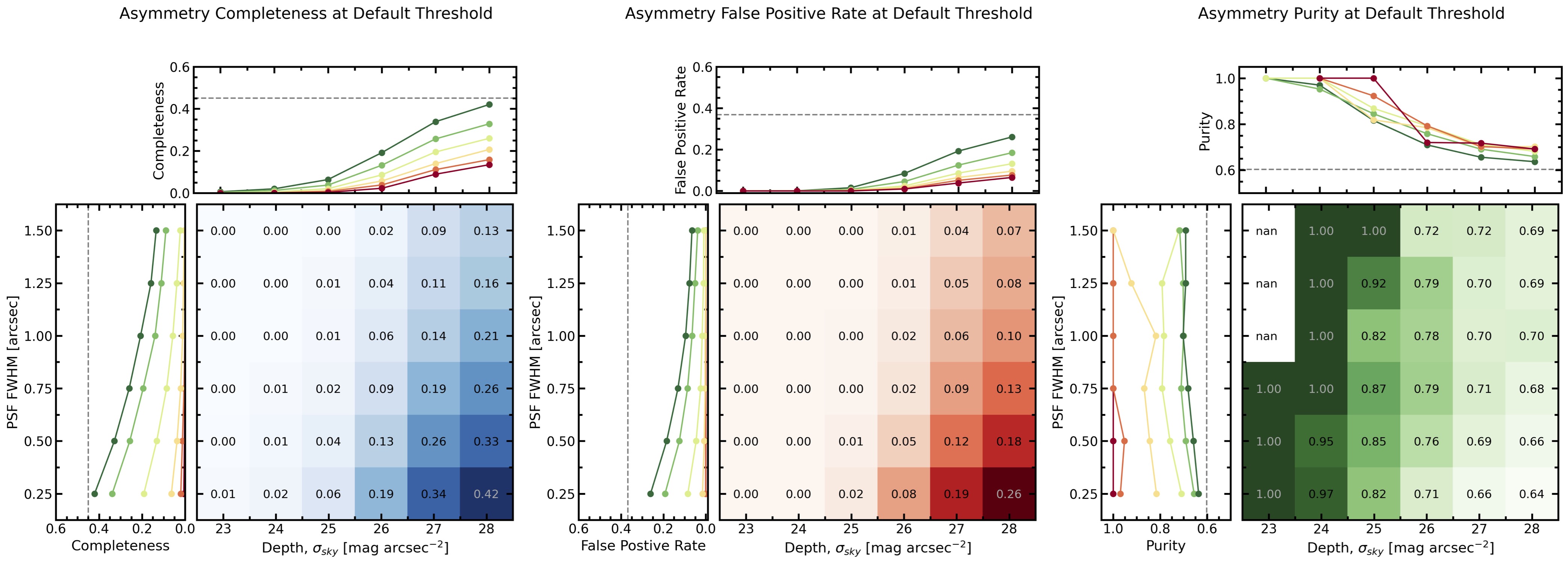}
        \caption{The completeness, false positive rate, and purity of the merger sample, as computed using asymmetry with the threshold $A>0.35$. This figure is composed of three subfigures, one for each the completeness, false positive rate, and purity. Each subfigure contains three panels. The bottom right panel is the most important; the blue/red/green colour gradient shows qualitatively the trend of completeness/false positive rate/purity as a function of both depth and PSF blurring, with specific completeness/false positive rate/purity values at each image quality reported in the corresponding cell. The top panel shows the relationship between completeness/false positive rate/purity and depth, with each line representing a different PSF FWHM. The lines are coloured from red to yellow to green with red indicating the worst image quality (highest PSF FWHM) and green indicating best image quality (lowest PSF FWHM). The left panel shows the relationship between completeness/false positive rate/purity and resolution, with each line representing a different depth. These lines are also coloured from red to yellow to green with red indicating the worst image quality (lowest depth) and green indicating best image quality (high depth). Similarly, the gray dashed lines indicate the performance in the ideal imaging. ``nan" values of purity refer to undefined values where there is a division by zero. Values of exactly 1.00 occur in cases where the completeness is between 0 and 0.005 and the false positive rate is precisely zero.}
        \label{ACompleteness}
    \end{figure*}

In Figure \ref{ACompleteness}, we present the completeness, false positive rate, and purity of the merger sample using the threshold $A>0.35$ to identify mergers as a function of depth and seeing in three subfigures. Each subfigure contains three panels. The main panel in the bottom right has depth and PSF FWHM on the x- and y-axes, respectively. The orientation of the axes is such that the worst image quality is in the top left corner and the best image quality is in the bottom right corner. At each combination of depth and PSF FWHM, the completeness/false positive rate/purity at that image quality is shown both qualitatively as a colour gradient (white is lowest and dark blue/red/green is highest) and quantitatively as a fraction between 0 and 1. The role of this panel is to showcase the qualitative trends of completeness/false positive rate/purity as a function of image quality and operate as a look-up reference table for other astronomers seeking an estimate of the completeness, false positive rate and/or purity of their merger sample, given their quality of imaging (see Section \ref{interpretation}). Indeed, all of the information of each subfigure is contained within the main panel (bottom right panel). However, the adjacent panels are added to demonstrate the individual trends of completeness/false positive rate/purity with PSF blurring and depth. The lines are coloured from red to green with red representing lower quality imaging and green representing higher quality imaging.

At the lowest quality of imaging tested (23 mag arcsec$^{-2}$ sky noise, 1.5 arcsec PSF FWHM) precisely zero mergers and non-mergers have asymmetries large enough to be classified as mergers by the threshold $A>0.35$. The blue and red colour gradients in Figure \ref{ACompleteness} show that as image quality improves, in both depth and seeing, the completeness (and false positive rate) increases. The highest completeness achieved is 42\% at the highest quality imaging considered (28 mag arcsec$^{-2}$ sky noise, 0.25 arcsec PSF FWHM). In contrast, the false positive rate at the highest quality imaging is 26\%.


Peculiarly, as image quality improves, the false positive rate increases more than the completeness. As a result, purity decreases as seeing and depth improve, independently, though more strongly in the case of depth. The purity computed from the $A>0.35$ method reaches a minimum of 64\% at the highest quality of imaging tested.



\subsubsection{Outer Asymmetry}

    \begin{figure*}
    \centering
    \includegraphics[width=1\linewidth]{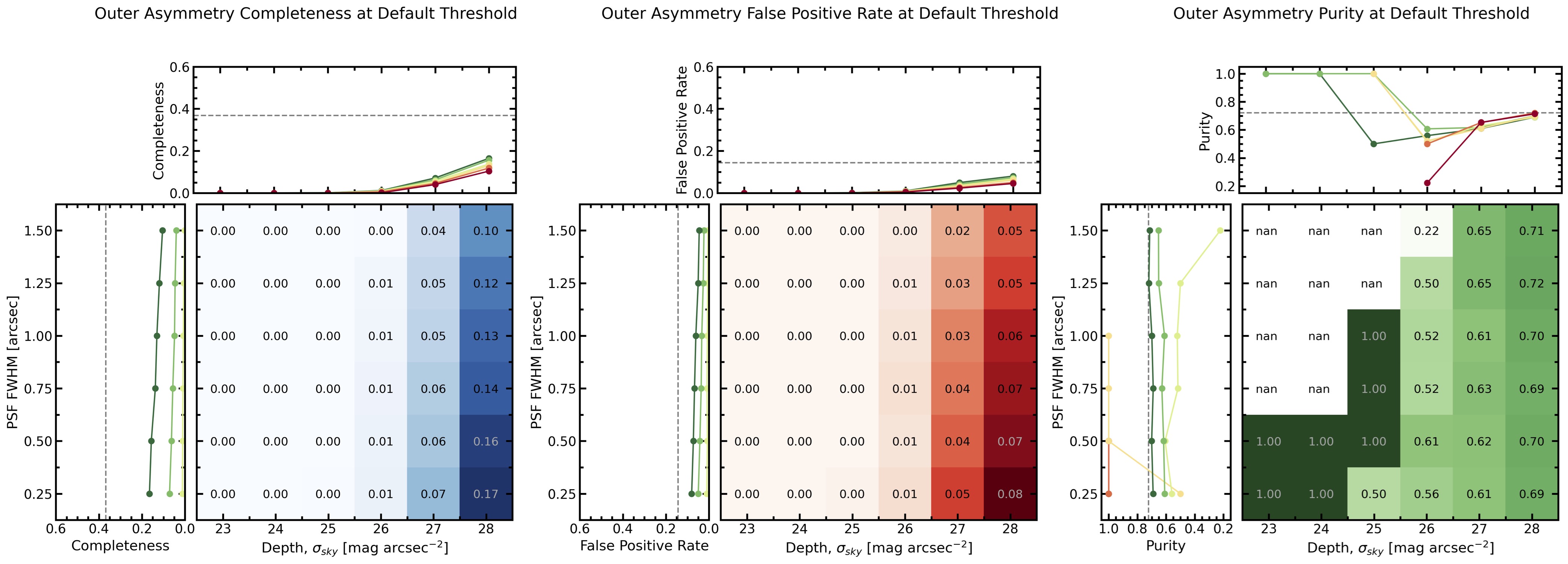}
        \caption{The same as in Figure \ref{ACompleteness}, but now considering the completeness, false positive rate, and purity of the merger sample, as computed using outer asymmetry with a threshold $A_O>0.6$.}
        \label{AoCompleteness}
    \end{figure*}

In Figure \ref{AoCompleteness}, we present the completeness, false positive rate, and purity of the merger sample recovered using the threshold $A_O>0.6$ as a function of depth and PSF blurring. Overall, completeness, false positive rate and purity of the outer asymmetry method all have a strong positive trend with depth, and a present but weak trend with seeing. At poor depth and seeing, precisely zero mergers and non-merger controls are detected as mergers, performing equally as poorly as standard asymmetry. As depth and seeing improve, the completeness, false positive rate and purity all increase (ignoring the anomalous 100\% purities resulting from poor sampling). In the highest quality imaging, the completeness reaches a maximum of 17\%, representing -25\% decrease in completeness compared to standard asymmetry. The false positive rate is also maximized in the highest quality imaging, at 8\%, which translates to a purity of 69\%, only about a +5\% improvement over asymmetry.

There are two key differences between the results for asymmetry and outer asymmetry. The first is that the outer asymmetry completeness and purity correlates more strongly with depth than seeing, whereas regular asymmetry showed similar trends with depth and seeing. Since outer asymmetry removes the central region of the galaxy from the asymmetry calculation, more statistical weight is being given to the faint exterior regions of the galaxy. As depth increases, the likelihood of a faint asymmetric feature being observed above the noise of the image increases, which makes deep imaging vital to identifying mergers with outer asymmetry. However, it is important to note that even in the deepest imaging, there is still a trend of increasing completeness with decreasing PSF blurring (see dark green line of left panel). This indicates that high resolution imaging is still beneficial for identifying some of the faint asymmetric features found in the outer regions of the mergers. 

The second key distinction in the outer asymmetry trends when compared to regular asymmetry is that the purity of outer asymmetry is low in shallower imaging and increases at higher depths. In the case of regular asymmetry, purity starts at higher values in shallow imaging and decreases in deeper imaging. As a consequence, regular asymmetry outperforms outer asymmetry at most image qualities, except in the deepest imaging tested here. We therefore find that asymmetry will generally identify purer samples of mergers, unless very deep imaging ($\sigma_\text{sky} \geq 27$ mag arcsec$^{-2}$) is used. 

\subsubsection{Shape Asymmetry}

    \begin{figure*}
    \centering
    \includegraphics[width=1\linewidth]{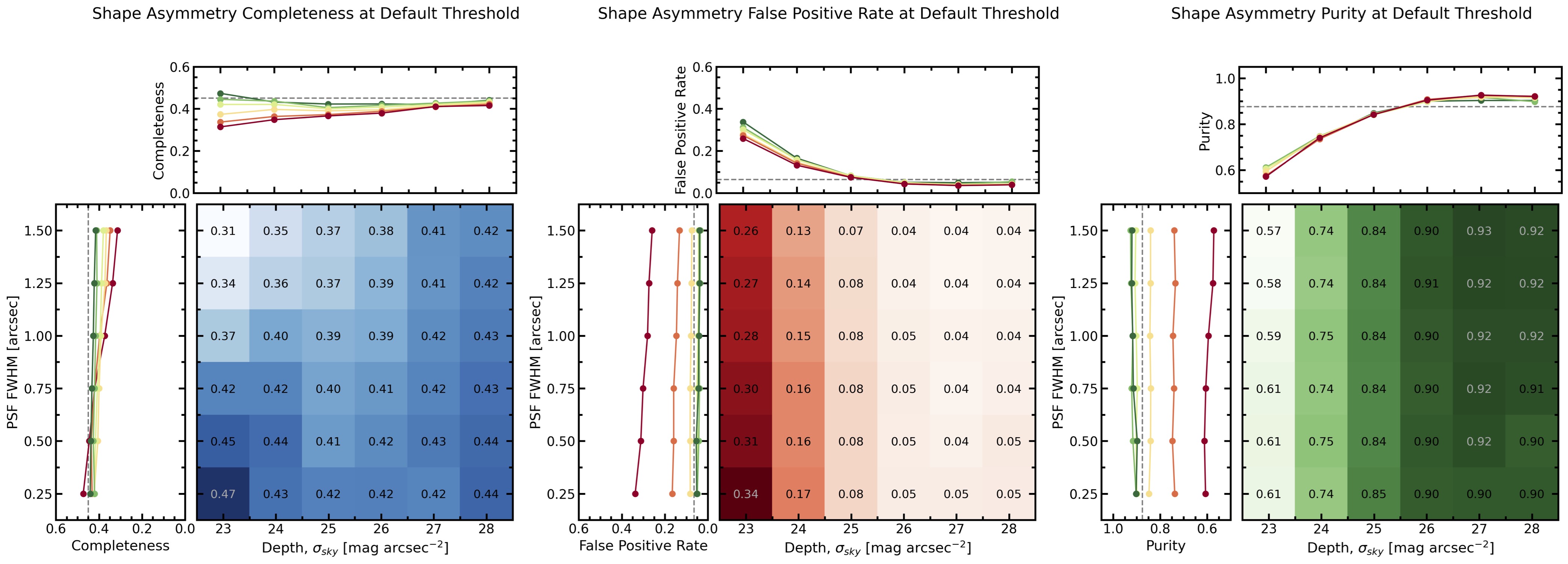}
        \caption{The same as in Figure \ref{ACompleteness}, but now considering the completeness, false positive rate, and purity of the merger sample, as computed using shape asymmetry with a threshold $A_S>0.4$.}
        \label{AsCompleteness}
    \end{figure*}

In Figure \ref{AsCompleteness}, we present the completeness of the merger sample recovered using the threshold $A_S>0.4$ as a function of depth and PSF blurring. Broadly speaking, shape asymmetry recovers more mergers than its light-weighted asymmetry counterparts $A$ and $A_O$; shape asymmetry achieves a completeness of 31\% in the lowest quality imaging when both $A$ and $A_O$ detected none. However, at the best image quality tested, shape asymmetry has a completeness of 44\%, +2\% better than asymmetry, but -18\% worse than outer asymmetry. This indicates that shape asymmetry produces samples that are always more complete than asymmetry, but at higher depths and resolutions, outer asymmetry identifies more mergers than shape asymmetry. The largest benefit of using shape asymmetry instead of the light-weighted asymmetry measurements is that shape asymmetry completeness is much more stable across image qualities (no strong trend with depth or resolution) and that it achieves a higher purity in high quality imaging.

It is worth noting that the stability of merger completeness with varying depths does not necessarily imply that depth has no effect on the measurement of shape asymmetry. Indeed, increasing the depth of the imaging does often change the measured shape asymmetry. In some cases, increasing the depth of the imaging reveals asymmetric, low surface brightness features previously hidden by sky noise thus increasing the shape asymmetry, possibly allowing a new merger detection to occur. In other cases, increasing the depth reveals a symmetric, diffuse stellar halo that may regularize the shape of a merger that had asymmetric features in the brightest parts of the galaxy. In such cases, it is possible that the increased depth can actually inhibit merger detection for a galaxy that may have been detected in noisier imaging. We thus interpret the relatively weak trend in completeness with depth as an approximately equal trade off between some mergers becoming more asymmetric and others becoming less asymmetric with increased depth, rather than depth having no effect on shape asymmetry merger detection. 

The shape asymmetry false positive rate is highest in shallow imaging, and decreases substantially as depth improves. The false positives are caused by only the brightest parts of the galaxy being visible above the noise. The shape of the galaxy is therefore being driven by bright and naturally occurring asymmetric features such as spiral arms and bursty star formation. This is further supported by the trend with seeing in the worst depth column; as the resolution of the shallow imaging increases, small asymmetric features are resolved and thus more false positives are detected. Indeed, in shallow imaging, the false positive rate is nearly as high as the completeness, leading to a purity of only 57\% in the lowest quality imaging.

Since the completeness remains roughly constant with depth and false positive rate decreases with depth, purity increases as depth increases. However, since the completeness and false positive rate trend proportionally with seeing, purity does not change much as a function of seeing. In the highest quality imaging tested, the purity achieved by shape asymmetry is 90\%.

\subsubsection{Gini-M$_{20}$ Merger Statistic}

    \begin{figure*}
    \centering
    \includegraphics[width=1\linewidth]{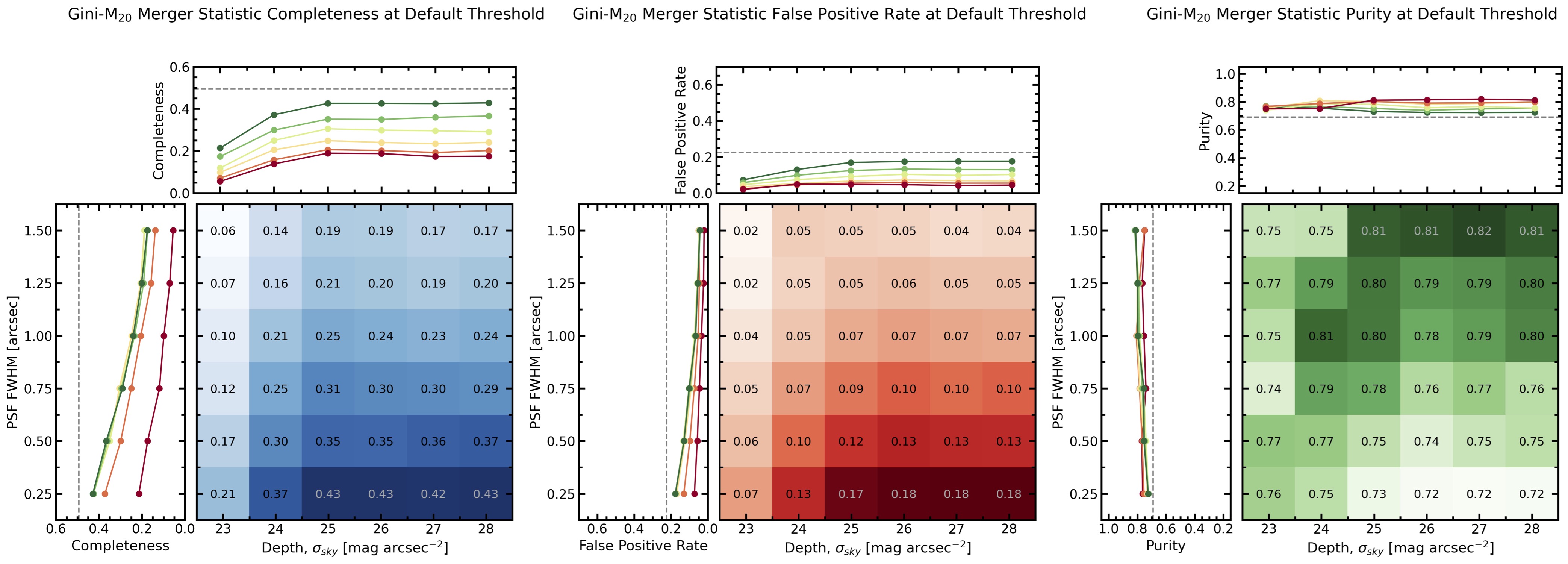}
        \caption{The same as in Figure \ref{ACompleteness}, but now considering the completeness, false positive rate, and purity of the merger sample, as computed using the Gini-M$_{20}$ merger statistic with a threshold $S(G,M_{20})>0$.}
        \label{GM20Completeness}
    \end{figure*}

    In Figure \ref{GM20Completeness}, we present the completeness, false positive rate, and purity of the merger sample recovered using the threshold $S(G,M_{20})>0$ as a function of depth and PSF blurring. The blue gradient shows that completeness increases with deeper imaging and better seeing, albeit with a stronger trend with seeing than depth. In the worst quality imaging, the completeness is 6\%, and in the highest quality imaging, the completeness is 43\%. The red gradient in the central subfigure shows a remarkably similar trend for the false positive rate as a function of image quality. The false positive rate is systematically lower than the completeness by a factor of 3-4 with the lowest (2\%) occurring in the worst quality imaging and the highest (18\%) occurring in the best quality imaging. Since the completeness and false positive rates share qualitatively similar trends with image quality, purity is relatively stable across all tested image qualities. In the lowest quality imaging, the purity is 75\% and in the highest quality imaging, the purity is 72\%. There is a slight decrease in purity as seeing improves and thus the highest purity (82\%) occurs at an image quality of 1.5 arcsec PSF FWHM and 27 mag arcsec$^{-2}$ sky noise.

    The completeness (and false positive rate) trend with seeing is the result of both Gini and M$_{20}$ decreasing due to blurry imaging. Gini will systematically decrease in worse seeing because blurring the image distributes the same amount of flux over a larger number of pixels. M$_{20}$ decreases because it is sensitive to the spatial variance and precise location of the brightest pixels containing 20\% of the flux which becomes regularized in the presence of PSF blurring. 

    The completeness (and false positive rate) trend with depth is caused by a similar effect as the false positive rate trend for shape asymmetry discussed in the previous subsection. In shallow imaging when sky noise is high, the extent of the galaxy decreases as faint features become dominated by sky noise. When fewer pixels belong to the galaxy and the total flux contribution is lower, fewer pixels are included in the Gini and M$_{20}$ calculations, which tends to increase the measurements (largely due to decreasing the denominators). Once sky noise has been reduced to include most of the possible galaxy flux, increasing depth to detect and include low surface brightness features does not affect the measurement of Gini or M$_{20}$ significantly (recall M$_\text{tot}$ weights galaxy pixels by flux) causing the trend with depth to flatten at $\sigma_\text{sky} \geq 25$ mag arcsec$^{-2}$.


    \subsubsection{Linear Discriminant Analysis}

\begin{figure*}
    \centering
    \includegraphics[width=1\linewidth]{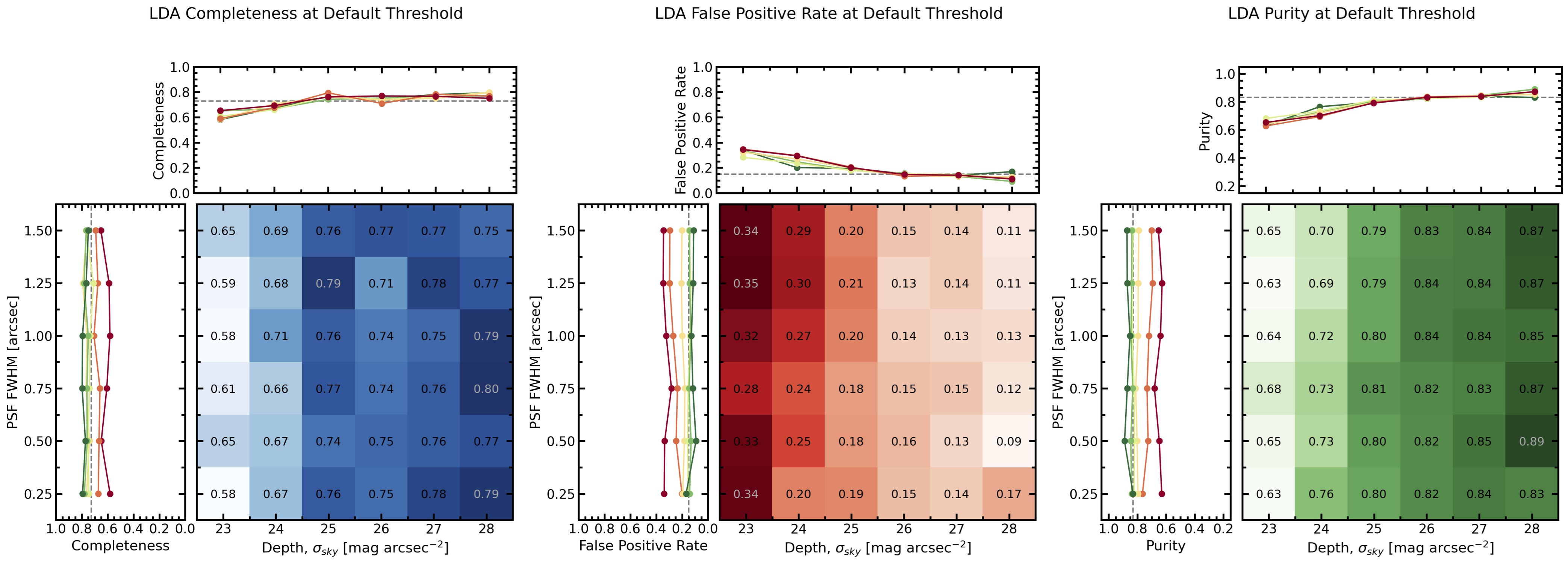}
        \caption{The same as in Figure \ref{ACompleteness}, but now considering the completeness, false positive rate, and purity of the merger sample, as computed using the LDA method with a threshold P$_\text{LDA}(\text{merger})>0.5$.}
        \label{LDACompleteness}
    \end{figure*}

    In Figure \ref{LDACompleteness}, we present the completeness of the merger sample recovered using the threshold P$_\text{LDA}(\text{merger})>0.5$ as a function of depth and PSF blurring. Overall, when the individual non-parametric morphology statistics are combined together with LDA, a substantially higher completeness is achieved at all image qualities (compared with individual statistics), while still maintaining fairly high purity. LDA completeness increases with depth but has no significant trend with seeing. In the lowest quality imaging, the LDA method achieves a completeness of 65\% and increases to 79\% in the highest quality imaging tested. The false positive rate has the opposite trend with depth and is also unaffected by changing seeing. In the lowest quality imaging, the false positive rate is 34\%, but decreases to 17\% in the highest quality imaging tested. Since the completeness increases and the false positive rate decreases with increasing depth, the purity of the sample steadily increases as depth improves. At the lowest quality of imaging, the purity is 65\% and at the highest quality imaging, the purity is 83\%.

    \begin{figure}
    \centering
    \includegraphics[width=1\linewidth]{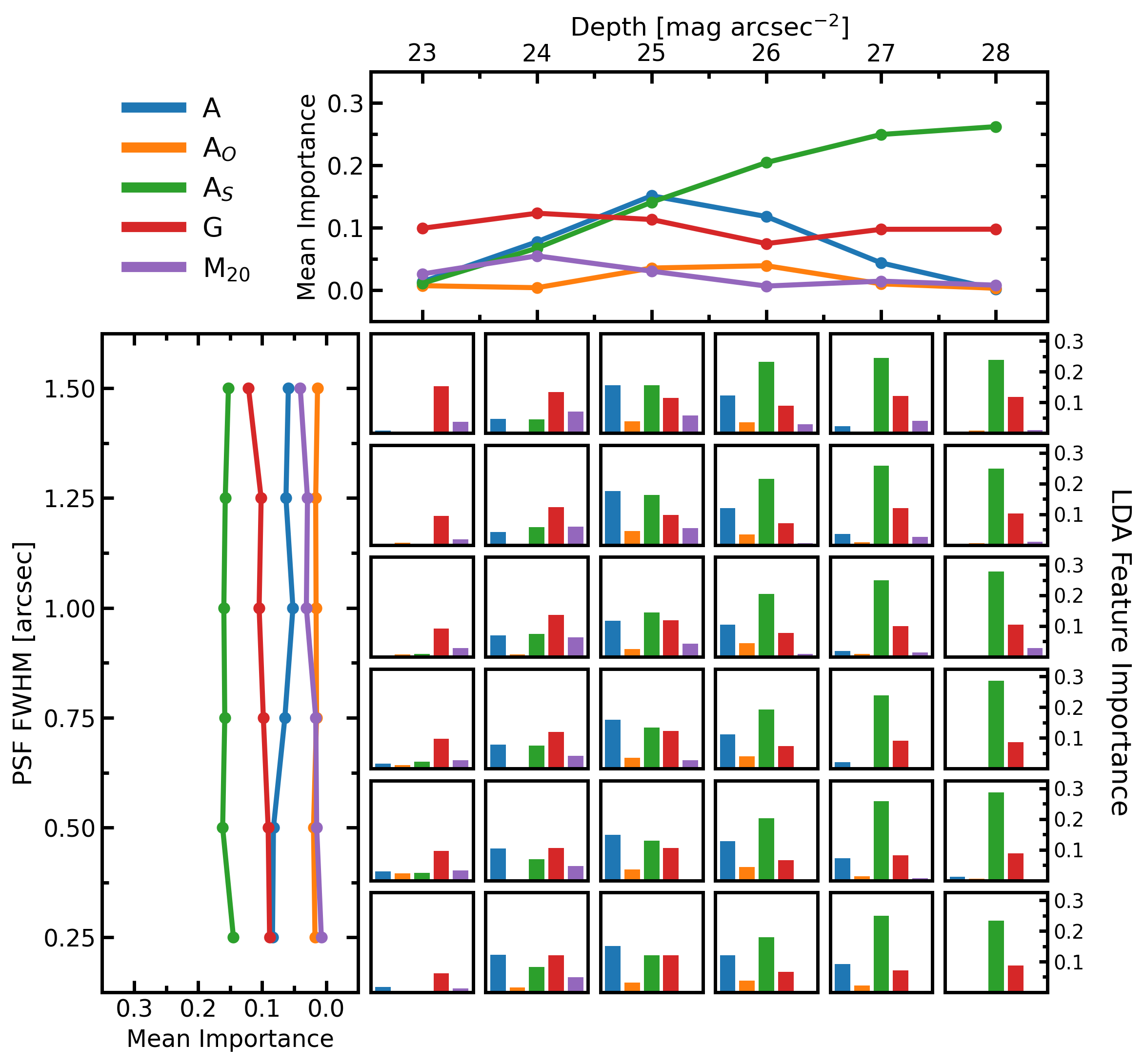}
        \caption{The feature importances of the LDA models presented as a function of image quality. Feature importance is computed as the drop in classifier accuracy when a feature of the test input is randomized. In the middle are 36 bar plots for each combination of depth and PSF FWHM. The colour of each bar represents the input statistic according to the legend in the top left of the figure and the height of the bars represents the importance of the statistic to the classifier accuracy at that image quality. The top panel shows the mean importance (at fixed depth) of each statistic as a function of depth and the left panel shows the mean importance (at fixed PSF FWHM) as a function of PSF FWHM.}
        \label{LDAImportance}
    \end{figure}

    To further inspect the LDA classification method, we compute the importance of the input features to the accuracy of merger classification. The accuracy is measured as the ratio of the number of correctly classified mergers and non-mergers to the total number of images classified ($\frac{N_\text{correct mergers} + N_\text{correct non-mergers}}{N_\text{total mergers} + N_\text{total non-mergers}}$). The importance of an input feature is then computed by randomizing one input column of the test set (i.e. no relationship between one of the morphology statistics and the merger/non-merger class) and measuring how much the classification accuracy of the test set drops. For a better statistical representation of feature importance, this is repeated ten times for each of the input features and the mean decrease in accuracy if a feature is randomized is taken to be the importance of that feature. 
    
    In Figure \ref{LDAImportance} we present the LDA feature importances as a function of image quality in the same format as Figure \ref{LDACompleteness}. The top panel of Figure \ref{LDAImportance} shows that the importance of individual features to the accuracy of the classifier varies with depth. In particular, shape asymmetry shows a clear increasing trend of importance as imaging gets deeper. Conversely, the asymmetry statistic increases as depth increases until at intermediate depths it turns over and becomes less important in deeper imaging. Focusing on the feature importance for the lowest quality imaging (top left bar plot), we find that the Gini and M$_{20}$ statistics have the greatest contribution to the LDA accuracy. We reason that this is because the Gini-M$_{20}$ statistics were combined together with higher redshift galaxies in mind \citep[see][]{Lotz04} where PSF blurring and depth pose a greater challenge for merger identification than at lower redshifts \citep{Ferreira18}. At intermediate image qualities all statistics contribute to the accuracy of the LDA classifier. Finally, in the highest quality imaging shape asymmetry is by far the most important feature for post-merger accuracy, with the Gini statistic also contributing. $A$, $A_O$, and M$_{20}$ do not contribute significantly to the LDA accuracy when the galaxies are embedded in deep imaging.

    One benefit of LDA is that the final output can be computed using a linear combination of normalized inputs, with coefficients output by the LDA itself. We present the LDA coefficients in a figure and as a reference table in Appendix \ref{LDAcoef}. Such a table allows anyone to use the LDA algorithms developed here to classify mergers and non-mergers in any dataset similar to one of the 36 image qualities presented here (see Section \ref{interpretation}).

    \subsubsection{Random Forest Classifier}

    \begin{figure*}
    \centering
    \includegraphics[width=1\linewidth]{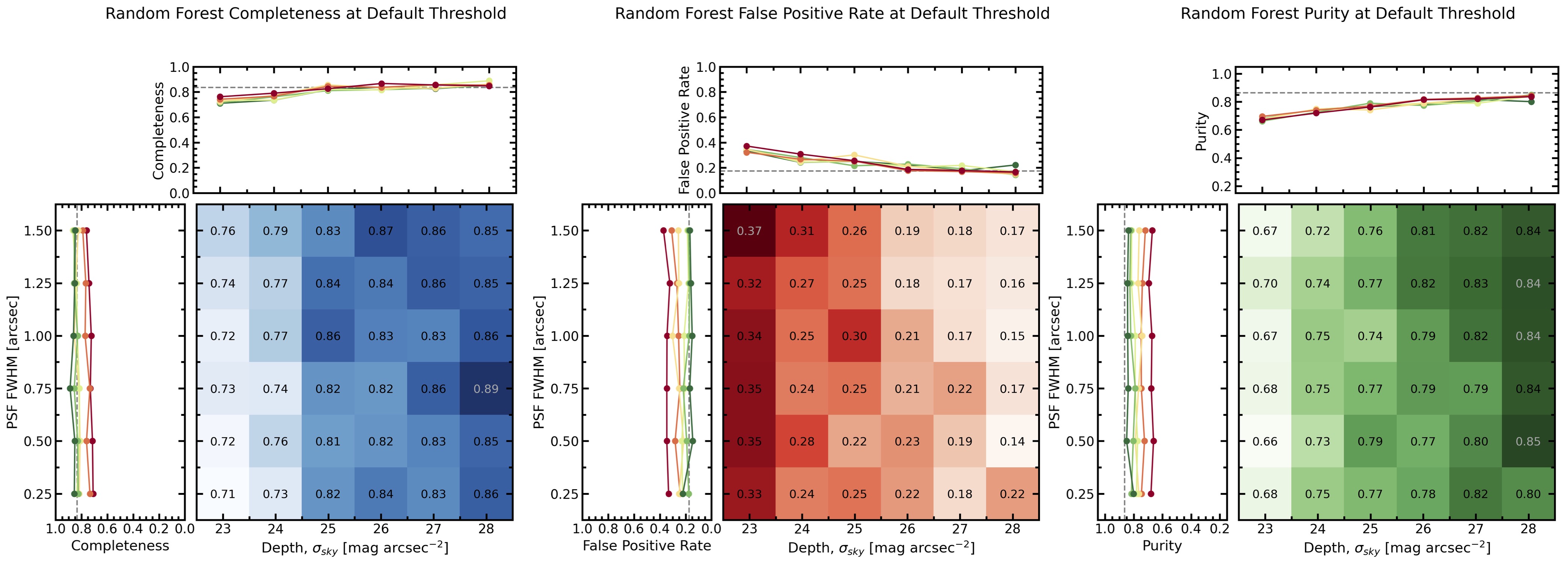}
        \caption{The same as in Figure \ref{ACompleteness}, but now considering the completeness, false positive rate, and purity of the merger sample, as computed using the random forest method with a threshold P$_\text{RF}(\text{merger})>0.5$.}
        \label{RFCompleteness}
    \end{figure*}

    In Figure \ref{RFCompleteness}, we present the completeness of the merger sample recovered using the threshold P$_\text{RF}(\text{merger})>0.5$ as a function of depth and PSF blurring. Broadly speaking, the random forest completeness, false positive rate, and purity trends are qualitatively the same as the LDA. However, when combining the individual non-parametric morphology with a random forest rather than LDA, a marginally higher completeness and purity is achieved at all image qualities. In the lowest quality imaging, the random forest achieves a completeness of 76\% (+11\% improvement over LDA) and increases to 86\% (+7\% improvement over LDA) in the highest quality imaging tested. However, the false positive rate is also higher than LDA in many (but not all) tested image qualities. In the lowest quality imaging, the false positive rate is 37\% (+3\% higher than LDA) and in the highest quality imaging tested, the false positive rate is 22\% (+5\% higher than LDA). The higher false positive rates in the random forest method causes a decrease in purity. When compared to the purity of LDA at the same image qualities, the random forest has lower purity than LDA for 24/36 image qualities, and lower purity for 23/24 image qualities with $\sigma_\text{sky} \geq 25$ mag arcsec$^{-2}$. Thus we conclude that while the random forest achieves higher completeness than LDA at all image qualities, the random forest produces higher purity in shallow imaging and LDA produces higher purity in deeper imaging.

    \begin{figure}
    \centering
    \includegraphics[width=1\linewidth]{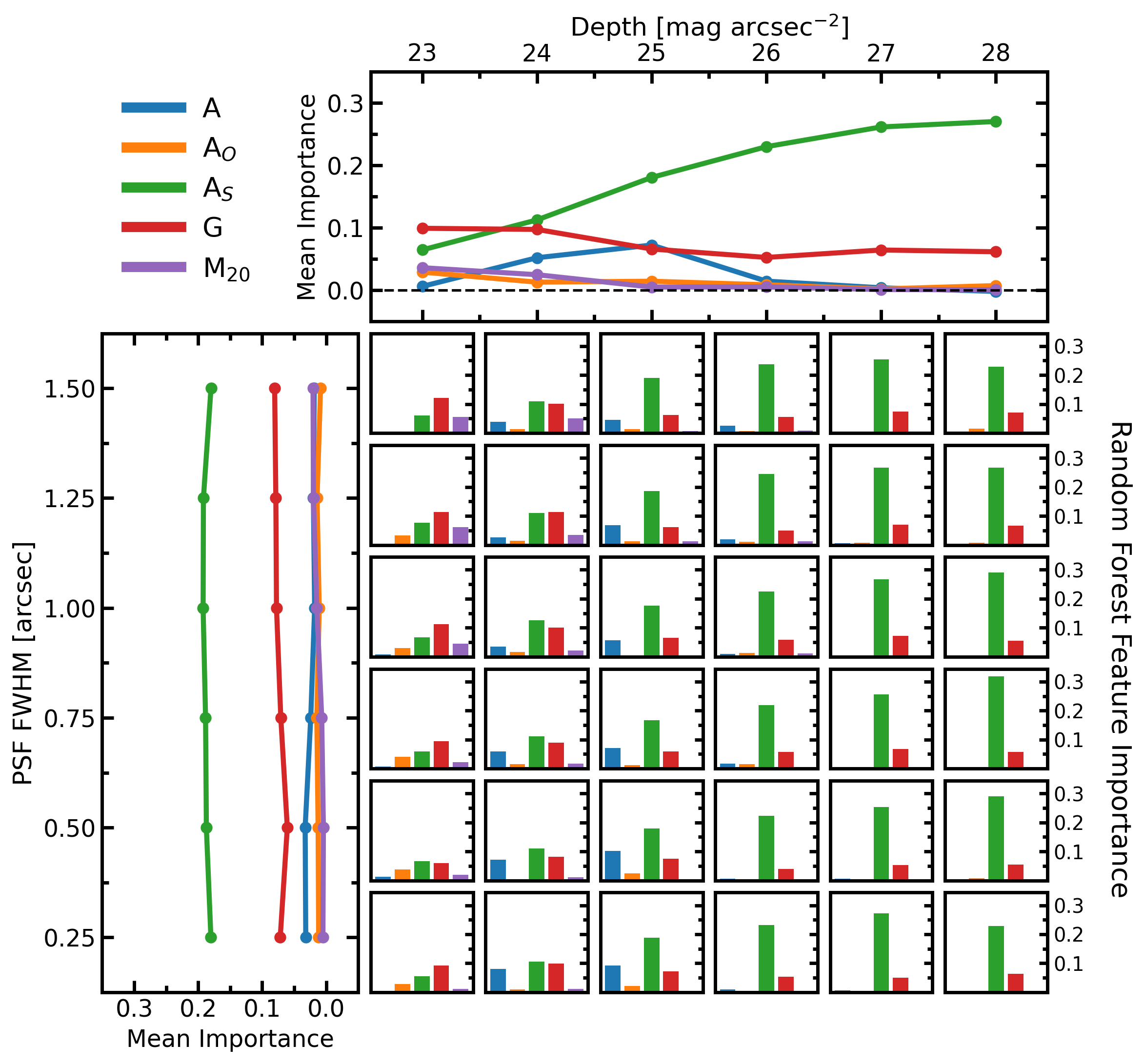}
        \caption{The same as in Figure \ref{LDAImportance}, but now considering the individual importance the non-parametric morphology statistics to the accuracy of the random forests.}
        \label{RFImportance}
    \end{figure}

    As was done for the LDA method, we further inspect the random forest classification method by computing the importance of the input features to the accuracy of merger classification, presented in Figure \ref{RFImportance}. The feature importance trends observed for the random forest models are very similar to those of the LDA; shape asymmetry increases as depth increases, asymmetry increases with depth but turns over and becomes less important in deep imaging, and Gini is relatively constant with depth but always contributing significantly to the classifier accuracy. However, there are several notable differences from the LDA feature importances. In the lowest quality imaging, Gini and M$_{20}$ are supported in the random forest by a significant contribution from shape asymmetry and at the lowest depth but higher resolutions, outer asymmetry becomes relevant too. Importance is shared across the features at intermediate image qualities, but not at all at depths greater than $\sigma_\text{sky} = 25$ mag arcsec$^{-2}$. At depths of $\sigma_\text{sky} \geq 25$ mag arcsec$^{-2}$, only asymmetry and Gini contribute significantly to the accuracy of the classifier. In this way, it seems that the random forest and LDA methods reach the same conclusion, shape asymmetry and Gini are most significant for merger classification in deep imaging, but the random forest converges to this solution faster (i.e. at lower quality imaging).



\subsubsection{Threshold-Independent Assessment of All Methods}
\label{ROCResults}

\begin{figure*}
    \centering
    \includegraphics[width=1\linewidth]{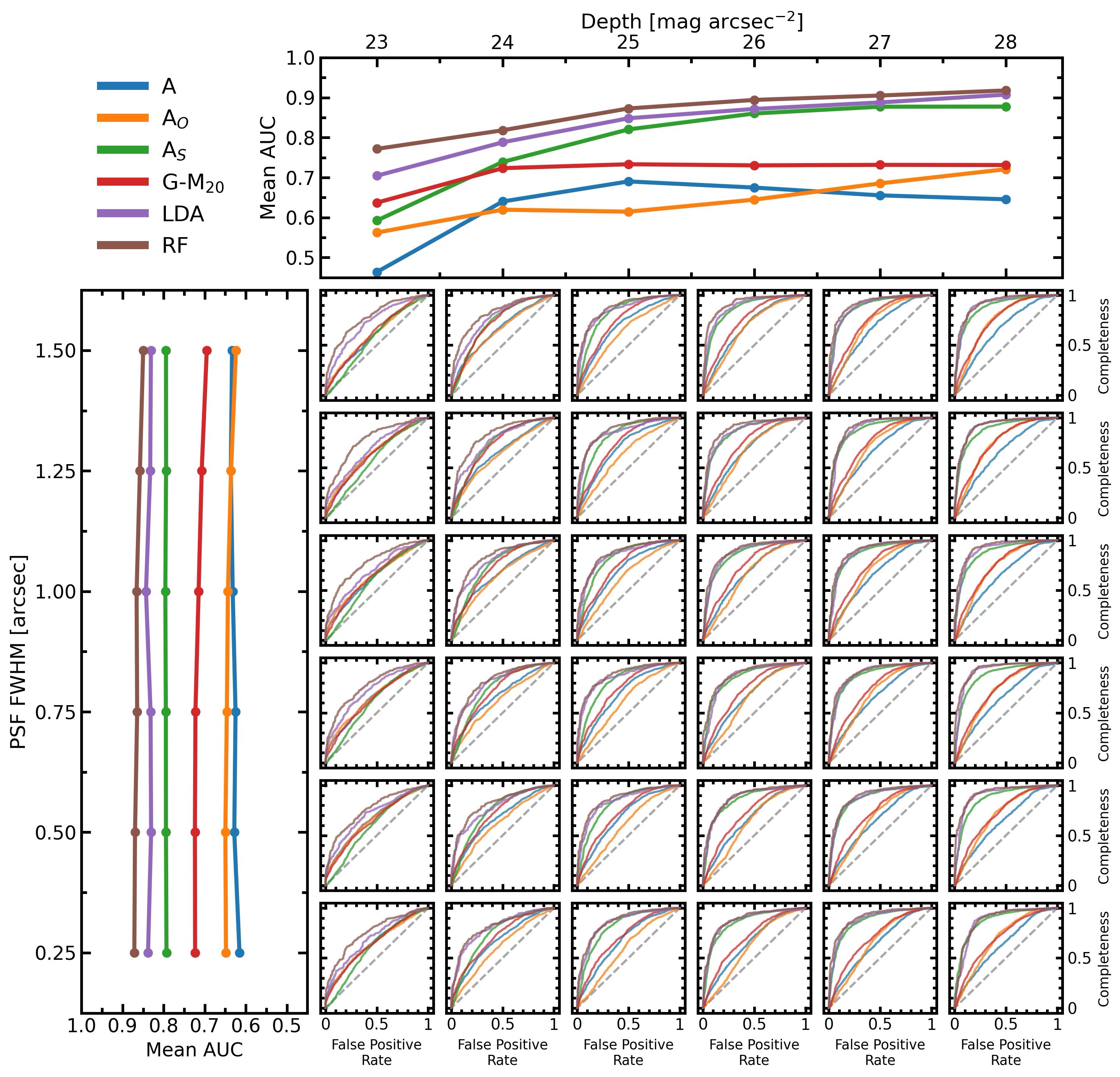}
        \caption{The ROC curves of the six merger identification methods as a function of image quality. In the central thirty six square panels are each methods ROC curve at the corresponding image quality, with the curves being coloured according to the legend in the top left of the figure. On each square panel, the gray dashed line represents the performance of a random classifier. The top panel shows the relationship between the mean area under the ROC curve (AUC) if each method at fixed depths and as a function of depth. The left panel shows the relationship between mean AUC (at fixed PSF FWHM) and resolution.}
        \label{ROCSummary}
    \end{figure*}

We have found that the individual non-parametric morphology statistics tend to have low completeness in poor image quality which increases as both seeing and depth improve. However, in the case of asymmetry, outer asymmetry, and the Gini-M$_{20}$ merger statistic, the false positive rate also increased as the image quality improved, often leading to impure merger samples, even in high quality imaging. The one exception to this was shape asymmetry, for which the completeness increased and the false positive rate \emph{decreased} as the image quality improved. We also found that when the non-parametric morphology statistics were combined together with classical machine learning methods, either LDA or random forest algorithms, the completeness and purity improved at all image qualities. Inspecting these methods in closer detail showed that in deep imaging, shape asymmetry had the most important impact on the LDA and random forest classifications. 

However, these results are true for only the default merger threshold (i.e. the threshold suggested in the literature). The LDA and random forest methods are weighing each of the non-parametric morphology statistics individually and effectively changing their merger thresholds in the process. In fact, in the case of the LDA method, it is \emph{just} recalibrating the merger threshold, albeit in higher dimensional feature space. 


Furthermore, adjusting the merger threshold from the default can be useful in specific circumstances. There is a natural trade-off between completeness and purity as the threshold is adjusted. Lower thresholds will be more inclusive of both mergers and non-mergers, likely increasing the completeness and false positive rate of the sample. High thresholds will be more exclusive, leading to incomplete but purer samples of mergers. Therefore, in cases where completeness is desired, a lower threshold will be more optimal, but in cases where purity is desired, a higher threshold will be more optimal. One compromise is to use the balance point threshold which equates the completeness to the one minus the false positive rate. We report the balance point threshold for each of the merger identification methods as a function of image quality in Appendix \ref{thresh}.

Without a specific threshold in mind, we present the ability of the six methods tested in this work to generate pure and complete samples using receiver operating characteristic (ROC) curves. ROC curves are assembled by computing the completeness and false positive rate at all possible thresholds and juxtaposing the two on the y- and x-axes, respectively. A perfect classifier makes an L-shape as the false positive rate can be zero with a completeness of 100\% and a random classifier would follow a diagonal line since at all thresholds the completeness and false positive rate are equal. The ability of a method to differentiate between mergers and non-mergers can be evaluated by measuring the area under its ROC curve (AUC). Accordingly, a perfect classifier would have AUC $=1$ and a random classifier would have an AUC $=0.5$.

In Figure \ref{ROCSummary}, we present the ROC curves for all six methods at each of the image qualities tested. The top panel shows the mean AUC (at a fixed depth) for each of the six merger identification methods as a function of depth. All methods demonstrate a trend towards higher AUC (i.e. better ability to differentiate mergers and non-mergers, regardless of threshold) in deeper imaging. The left panel shows the mean AUC (at fixed PSF FWHM) as a function of PSF FWHM. All of the merger identification methods tend towards higher AUC in higher resolution imaging, but the trend is much weaker than that observed for depth. Together, this demonstrates that deep imaging is more important for distinguishing post-mergers from non-mergers than high resolution. 

The coloured ROC curves in the central 36 panels demonstrate that all merger identification methods tested here are better than random classifiers (i.e. the ROC curves are all above the diagonal dashed line). In the lowest quality imaging the LDA and random forest methods produce ROC curves with a much larger AUC than all of the individual non-parametric morphology statistics. This further demonstrates the importance and power of combining the non-parametric morphology statistics together for merger classification. In deep imaging ($\sigma_\text{sky} \geq 26$ mag arcsec$^{-2}$), the shape asymmetry statistic becomes a better classifier, with an ROC curve rivalling and converging with those of the LDA and random forest methods. This is consistent with shape asymmetry becoming the most important input feature to the LDA and random forest models at these depths as seen in the analysis of LDA and random forest feature importances (see Figures \ref{LDAImportance} and \ref{RFImportance}). Though the LDA and RF models always achieve higher AUC than shape asymmetry alone, emphasizing the importance of including additional metrics in the training process.

\subsection{The Effect of Galaxy Properties}
\label{Results-props}

In the previous section, we have critically assessed the completeness of a sample of mergers degraded to various levels of image quality. However, it has been previously observed that other factors such as stellar mass, mass ratio, orbital and dynamic characteristics, and gas fraction can also affect the observability of a recent merger \citep{Bell06,Lotz10-gas,Lotz10-mu, Snyder19, Nevin19, McElroy22}. In this section, we break down the merger sample from the previous section into smaller groups based on the mass ratio of the merger (Section \ref{MassRatio}), the total stellar mass of the post-merger (Section \ref{StellarMass}), and the gas fraction of the post-merger (Section \ref{GasContent}). This will allow us to inspect how each of these properties affect a merger's likelihood of being detected by each of the merger identification methods and how these properties have contributed to the trends observed in Section \ref{Results-IQ}.

\subsubsection{Mass Ratio of Merger}
\label{MassRatio}

\begin{figure*}
    \centering
        \includegraphics[width=1\linewidth]{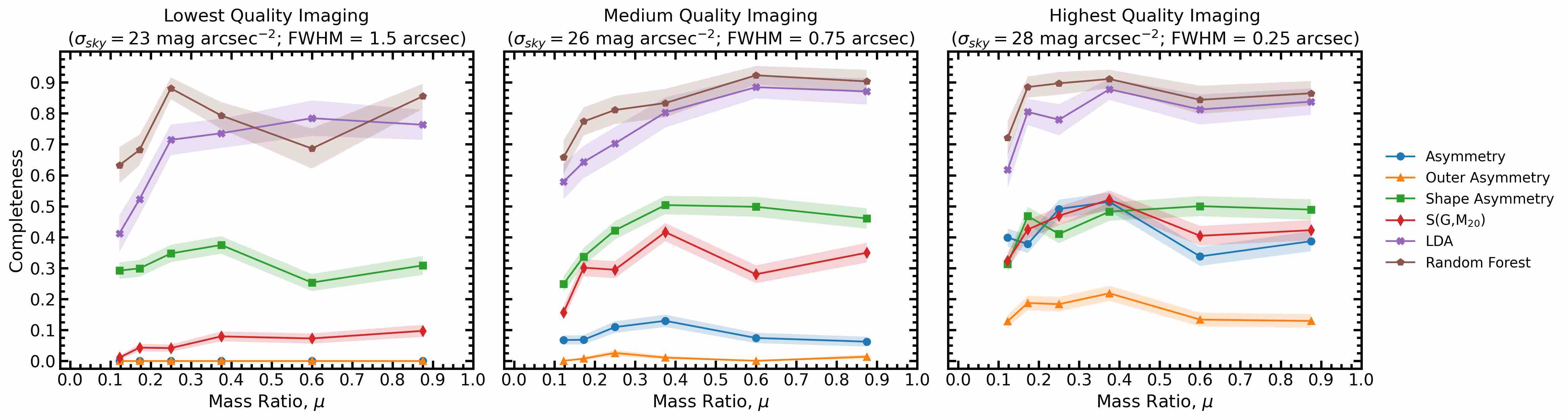}
        \caption{The completeness of the detected mergers using the six merger identification methods and their default merger detection thresholds binned by the mass ratio of the galaxy mergers. The completeness is calculated in an adaptive binning scheme that allows the size of the bins to change in order to maintain equal number of galaxy images ($\thicksim 280$) in each bin. Each panel shows the completeness trends for asymmetry (blue circles), outer asymmetry (orange triangles), shape asymmetry (green squares) and the Gini-M$_{20}$ merger statistic (red diamonds), LDA (purple crosses), and the random forest (brown pentagons) at a given image quality. The shaded regions around the points are representative of the error on the completeness values, computed using binomial counting statistics. In the left panel are the completeness trends for the lowest image quality tested (23 mag arcsec$^{-2}$ sky noise and 1.5 arcsec PSF FWHM), in the central panel are the completeness trends for an intermediate image quality (26 mag arcsec$^{-2}$ sky noise and 0.75 arcsec PSF FWHM), and in the right panel are the completeness trends for the highest quality imaging tested.}
        \label{Completeness-mu}
    \end{figure*}

    The mass ratio of the two progenitor galaxies can have a significant impact on the observed morphology and therefore, their ability to be detected as mergers. In general, galaxies with mass ratios closer to one will have more significant and longer lasting morphological disturbances \citep{Lotz10-mu, Casteels14, Ji14, Nevin19}. In Figure \ref{Completeness-mu}, we present the completeness of the detected merger sample achieved by each of the six merger identification methods in bins of mass ratio and at three different image qualities. An adaptive binning method is used such that the bins of mass ratio change in size in order to preserve a roughly equal number of galaxies in each bin. The shaded area around each curve corresponds to the error computed using binomial counting statistics. For the non-parametric morphology statistics the entire sample is used. In contrast, only the test set (30\% of the total sample not seen during training) is used to compute the completeness for the LDA and random forest methods. 
    

    In the left panel of Figure \ref{Completeness-mu}, the completeness of the recovered mergers is determined using each of the six methods applied to the lowest quality imaging (23 mag arcsec$^{-2}$ depth and 1.5 arcsec PSF FWHM). This image quality is very poor, worse than that of SDSS. Since no mergers are identified by either of the light-weighted asymmetry metrics (A and A$_O$) at this image quality, there is no trend between completeness and mass ratio to discuss. Shape asymmetry has a completeness of 25-35\% in each bin, but has no clear trend with mass ratio. We do not interpret this as shape asymmetry being equally good at detecting high and low mass ratio mergers in poor quality imaging. Rather, since the false positive rate is 26\% at this image quality, it is more accurate to say that shape asymmetry is not working effectively at this image quality causing mass ratio to be irrelevant to merger classification. However, the Gini-M$_{20}$ method has a completeness of $\thicksim 0\%$ in the lowest mass ratio bin and increases monotonically as mass ratio increases to a maximum of $\thicksim 10\%$ in the highest mass ratio bin. LDA also has a strong trend with mass ratio, achieving 41.2\% completeness in the lowest mass ratio bin and 76.3\% in the highest. Finally, the random forest has no clear trend with mass ratio. Regardless, the random forest still has a worse completeness in the lowest mass ratio bin (63.2\%) than in the highest mass ratio bin (85.5\%).

    In the centre panel of Figure \ref{Completeness-mu}, we show the completeness of the recovered mergers as determined using each of the six methods applied to an intermediate-quality imaging (26 mag arcsec$^{-2}$ depth and 0.75 arcsec PSF FWHM). This image quality is similar to many present-day (and forthcoming) ground based photometric surveys such as CFIS, the Kilo-degree Survey \citep{deJong13}, and the 1-year LSST depth \citep{Ivezic19}. Therefore, the completeness trends presented here are representative of biases in merger samples in recent and upcoming merger searches at this image quality. In this case, all of the metrics have positive trends with mass ratio, except for $A$ and $A_O$ which detect very few mergers and have no clear trend with mass ratio. Thus, merger samples assembled using these detection methods are generally biased towards higher mass ratio mergers. Furthermore, if we were only looking for major mergers, we would report a much higher completeness in Figures \ref{ACompleteness}-\ref{RFCompleteness}.
    

    In the right panel of Figure \ref{Completeness-mu}, we show the completeness of the recovered mergers as determined using each of the six methods applied to the highest quality imaging tested in this work (28 mag arcsec$^{-2}$ depth and 0.25 arcsec PSF FWHM). This image quality is comparable to the 10-year LSST depth, but with better seeing \citep{Laine18, Ivezic19, Brough20, Martin22}. Therefore, the trends presented here are representative of the best we may be able to accomplish with these merger identification methods in our current era of ground-based astronomy surveys. At this image quality, asymmetry, outer asymmetry and the Gini-M$_{20}$ merger statistic have overall higher completeness, consistent with the results in Section \ref{Results-IQ}). These three methods also have complicated trends with mass ratio; there is an increase in completeness from low mass ratio bins to intermediate mass ratio bins, only to turn over again at higher mass ratios. The other three methods, shape asymmetry, LDA and the random forest have similar completeness in the high mass ratio bins but significantly improved completeness in the lower mass ratio bins. Therefore, we conclude that a primary benefit of increasing the quality of imaging is to identify more low mass ratio mergers, which have been shown to contribute significantly to merger-triggered star-formation enhancements \citep{Hani2020, Bottrell23}. At this image quality, shape asymmetry, LDA and random forest methods can produce more complete and less biased merger samples than in the current state-of-the-art photometric surveys.

\subsubsection{Total Stellar Mass of Merger}
\label{StellarMass}

\begin{figure*}
    \centering
        \includegraphics[width=1\linewidth]{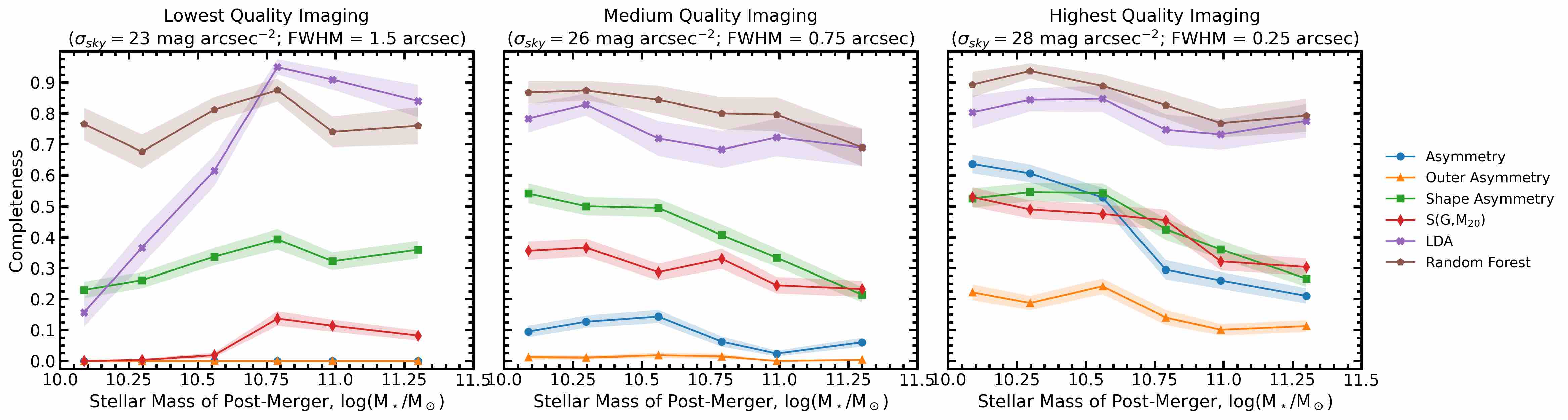}
        \caption{The same as in Figure \ref{Completeness-mu}, but now considering the completeness of the detected mergers in bins of stellar mass at three different image qualities.}
        \label{Completeness-Mstar}
    \end{figure*}

    Another important property of galaxies for measuring morphology and studying galaxy evolution in general is the total stellar mass of the system \citep{Casteels14, GO23}. In Figure \ref{Completeness-Mstar}, we present the completeness of the detected merger sample achieved by each of the six merger identification methods in bins of total stellar mass and at three different image qualities. Bins of stellar mass are adaptive in order to preserve a roughly equal number of galaxies in each bin. The shaded area around each curve corresponds to the error computed using binomial counting statistics.

    The left panel of Figure \ref{Completeness-Mstar} shows that in shallow imaging with poor seeing, several merger identification metrics are more complete at higher stellar masses than at lower stellar masses. This is likely because more massive galaxies are brighter and more likely to have asymmetric features visible above the noise. This trend is most evident in the LDA completeness. In the lowest stellar mass bin LDA has a completeness of only 15.1\%, which increases to a staggering 94.5\% in the second highest mass bin. The purity is 72\% and 62\% in each of those bins, respectively.

    The centre panel of Figure \ref{Completeness-Mstar} shows that all of the merger identification methods follow the opposite trend than that which was dominant in the lower quality imaging. Once a baseline quality of imaging is met, lower mass galaxies are actually more likely to be detected as mergers than higher galaxies. The same trend holds for the highest image quality, shown in the right panel of Figure \ref{Completeness-Mstar}. However, the LDA and random forest methods exhibit a relatively flat relationship between completeness and stellar mass indicating that in high quality imaging they can produce much less biased samples of mergers than any of the individual non-parametric morphology statistics. 
    

    It is at this point that we would like to remind the reader that these results do not necessarily imply that the high stellar masses themselves are causing galaxies to be less likely to be detected as a recent merger. Such a statement would require careful controlling of other factors that may affect merger detection (e.g. mass ratio of merger, gas fraction, and environment). Our results are instead a true manifestation of what the completeness may look like in given mass regimes, with realistic distributions of gas fractions, mass ratios and environments, according to the cosmological context provided by TNG100. For example, the high stellar mass galaxies generally have lower gas fractions which may be contributing to the low completeness. We investigate this possibility explicitly in the following subsection.

\subsubsection{Gas Content of Merger}
\label{GasContent}

\begin{figure*}
    \centering
        \includegraphics[width=1\linewidth]{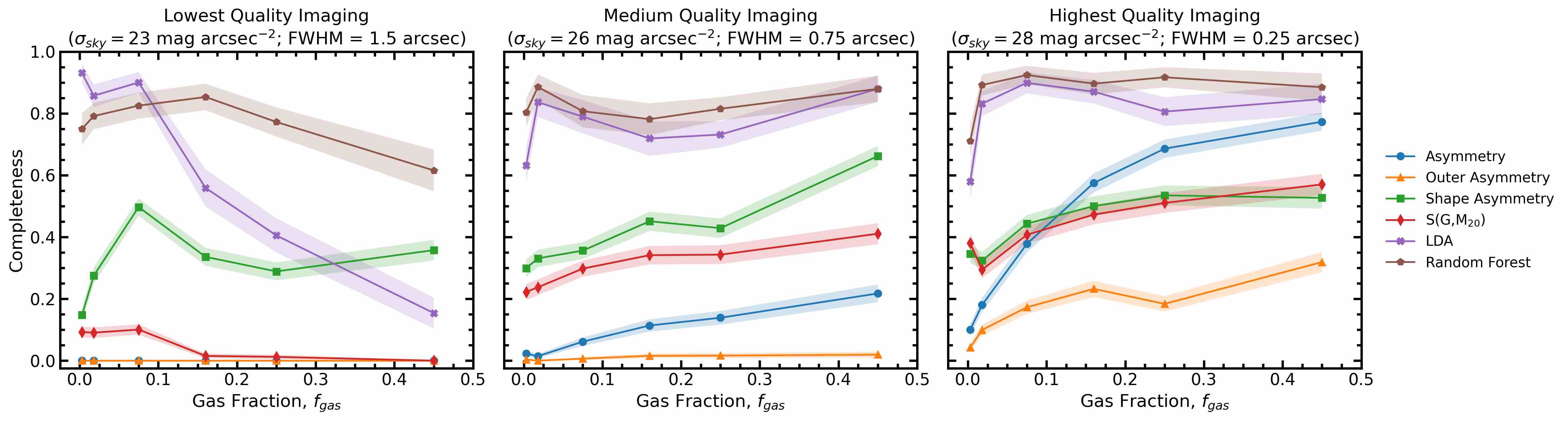}
        \caption{The same as in Figure \ref{Completeness-mu}, but now considering the completeness of the detected mergers in bins of gas fraction at three different image qualities.}
        \label{Completeness-fgas}
    \end{figure*}

    Another driving property of galaxy morphology is the gas fraction \citep{Bell06, Lotz10-gas}. 
    In Figure \ref{Completeness-fgas}, we present the completeness of the detected merger sample achieved by each of the six merger identification methods in bins of post-merger gas fraction and at three different image qualities. As in Figures \ref{Completeness-mu} and \ref{Completeness-Mstar}, the bins are adaptive in order to preserve a roughly equal number of galaxies in each bin and the shaded area around each curve corresponds to the error computed using binomial counting statistics.

    The left panel of Figure \ref{Completeness-fgas} shows that in the lowest quality image tested, several of the merger identification methods have completeness trends that are anti-correlated with gas fraction. Since gas fraction is expected to increase merger detection, we suspect that this trend is a manifestation of the correlation between low gas fractions and high mass galaxies, which were seen to have high completeness in Figure \ref{Completeness-Mstar}. This is supported by the fact that, like the trends in Figure \ref{Completeness-Mstar}, the trends in Figure \ref{Completeness-fgas} invert direction in intermediate quality imaging.

    The central and right panels of Figure \ref{Completeness-fgas} show that once a baseline image quality is met, the individual non-parametric morphology statistics are very sensitive to gas fraction, with higher gas fractions allowing for higher completeness in the post-merger sample. In the most extreme case, the asymmetry method applied to the highest quality imaging (see right panel of Figure \ref{Completeness-fgas}) has a completeness of only 10.0\% in the lowest gas fraction bin but a completeness of 77.3\% in the highest gas fraction bin. However, it is worth mentioning that the purity achieved by asymmetry is actually highest in the lowest two gas fraction bins where completeness is quite low, and as the completeness increases, the purity remains roughly constant between 65-70\%. Therefore, gas fraction may simply be fuelling bursty star formation, which is more common (or stronger) in mergers but not exclusive to recent mergers. 

    We would also like to emphasize that both the LDA and random forest methods in the intermediate and high quality imaging -- particularly the random forest in the highest image quality -- produce nearly unbiased samples of mergers, with respect to gas fraction. This represents a significant improvement over the individual non-parametric morphology statistics which are all biased towards identifying higher gas fraction post-mergers.

\subsection{The Effect of Orientation}
\label{Discussion-TNG}
\label{disc-ViewAngExp}

So far, we have shown that even in ideal imaging, the maximum completeness using any of the individual non-parametric morphology statistics is 55\% (see Figure \ref{OneIQ-dists}). Combining the non-parametric morphology statistics using a random forest allowed for 86\% completeness to be attained in ideal imaging. Moreover, we have shown that realistic imaging effects and various properties of the galaxies can lead to lower completeness. Even in the most favourable conditions for observing mergers ($\sigma_\text{sky} = 28$ mag arcsec$^{-2}$, $\text{FWHM} = 0.25$ arcsec, $\mu>0.3$, $M_\star<10^{11} M_\odot$, $f_\text{gas} > 0.2$) only 91.9\% of recent mergers are identified successfully by the random forest method. What about the remaining 8\% that are not detected? In this section, we explore the dependence of merger observability on the viewing angle from which the imaging is generated. We find that there may be an upper limit to the completeness of merger identification due to the orientation of the post-mergers and their telltale features. 

To test the effect of orientation on merger observability, we start by taking one galaxy from our post-merger sample and generate synthetic imaging following the methods discussed in Section \ref{SyntheticImages} for 648 viewing angles instead of four. The viewing angles are placed 10 degrees apart in inclination (from $-$90$^\circ$ to 90$^\circ$) and azimuth (from $0^\circ$ to 360$^\circ$) for a total of $36\times18 = 648$ viewing angles. Unlike the four viewing angles used in the main analysis of this work, the small differences in viewing angle are used to show the gradual change in observed morphology. For this test, the images are generated with an image quality similar to that of present day state-of-the-art photometric surveys (PSF FWHM of 0.75 arcsec and depth of 26 mag arcsec$^{-2}$). 

\begin{figure*}
    \centering
        \includegraphics[width=\linewidth]{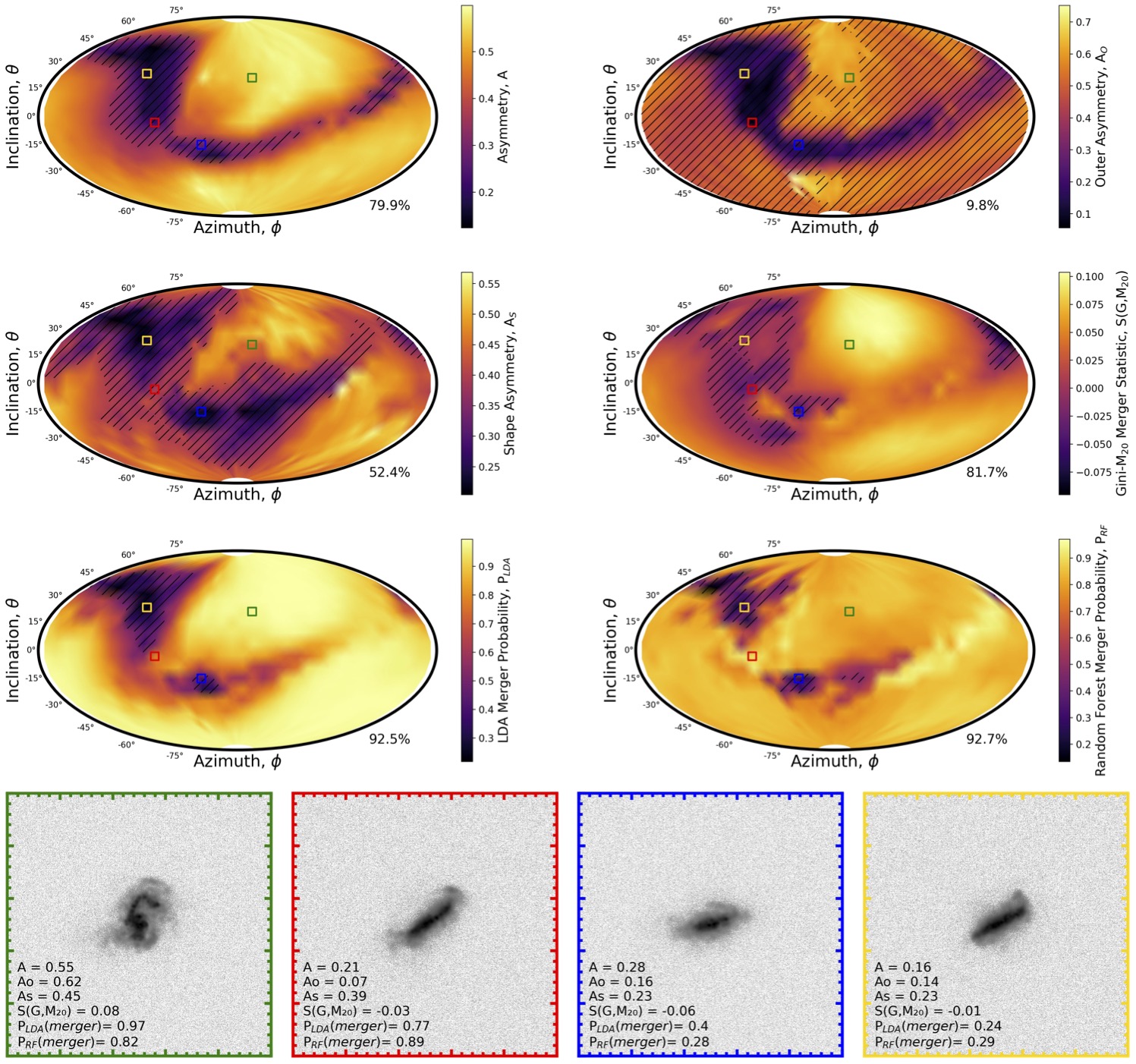}
        \caption{An example of the variation of the non-parametric morphology measurements and post-merger probabilities over all viewing angles for one TNG100 post-merger (snapshot 95, subhalo 474801). The measurements are made in 10$^\circ$ increments in azimuth and inclination. The maps are generated using an Aitoff projection of a sphere to 2-D and smoothed using Gouraud shading. The value of each statistic or merger probability at each viewing angle is indicated by the colour gradient and quantified by the colour bars on the right of each plot which range from the minimum to the maximum of each individual statistic. Viewing angles where the morphology statistic are below the default merger threshold are demarcated by black hatches and the fraction of viewing angles where the morphology statistic is above the default merger threshold is given in the bottom right of each panel. Coloured squares in each panel denote the viewing angles of the example images shown along the bottom with axes coloured accordingly.}
        \label{ViewAngExp}
    \end{figure*}

In Figure \ref{ViewAngExp} we present the measured non-parametric morphology statistics asymmetry, outer asymmetry, shape asymmetry, and the Gini-M$_{20}$ merger statistic and the post-merger probabilities computed by the LDA random forest methods at each of the 648 viewing angles considered for one example post-merger galaxy. This post-merger (snapshot 95, subhalo 474801) has a total stellar mass of 3.1$\times10^{10}M_\odot$, gas fraction of 0.37, and has undergone a merger with a mass ratio of 0.26 within the last snapshot of the simulation ($t_\text{postmerger} \lesssim 150$ Myr). The map in each panel of Figure \ref{ViewAngExp} shows a 2-D Aitoff projection of the morphology statistic or merger probability, coloured using Gouraud shading\footnote{Gouraud shading is built in to the \texttt{matplotlib pcolormesh} function and determines the colour of each grid point by linearly interpolating from the corners of each grid in the mesh.}. In each panel, the morphology (as quantified by each of the six metrics) is seen to vary considerably with viewing angle; the colour bars adjacent to each panel demonstrate that the maxima and minima reached at different viewing angles of this galaxy span nearly the entire possible dynamic range for the methods used. Accordingly, there are many viewing angles at which this bona fide merger is not above the default merger threshold and therefore would not be detected as a merger. The viewing angles for which the method does not detect this galaxy as a merger are indicated by black hatch marks. The fraction of viewing angles for which this galaxy would be detected as a merger is indicated in the bottom right corner of each panel. Since the viewing angles are evenly spaced in azimuth and inclination, angles at high inclinations subtend less solid angle than those at low inclination. When determining the fraction of viewing angles above the merger threshold, the viewing angles are weighted by $\cos(\theta)$ to account for the difference in solid angle.

For this example galaxy, the six merger identification methods have similar trends with viewing angle, succeeding and failing at similar orientations. Each of the panels show a dark band of viewing angles from which the merger is not detected (or is only marginally detected). Away from this band, most metrics are maximized and constant, with the exception of shape asymmetry which seems to vary from one viewing angle to the next. 

To inspect these trends, we have included the images from four key viewing angles along the bottom of Figure \ref{ViewAngExp}. The selected viewing angles are highlighted by green, red, blue and yellow boxes on the six panels with each coloured box corresponding to the colour of the images' axes. The green box shows a viewing angle for which all merger identification methods detect this galaxy as a merger. The corresponding image with green axes shows that indeed, the merger features of this galaxy are very clear. There are internal asymmetric disturbances, a bright extended asymmetric feature, as well as low surface brightness asymmetric features which distort the overall shape of the galaxy. The red box shows a viewing angle for which all of the individual non-parametric morphology statistics are unable to identify the merger, but the combination of the statistics through LDA and random forest allows the merger to be identified. At this viewing angle, merger features are less obvious; the disk has slightly distorted shape and there are low surface brightness features in the bottom left and top right that are not symmetric. It is worth noting that shape asymmetry (at the default threshold of $A_S > 0.4$) almost identifies this as a merger ($A_S = 0.39$ at this angle), but ultimately does not. Finally, the blue and yellow boxes are two examples for which \emph{none} of the six merger identification methods correctly identified this merger. However, who could blame them? At these viewing angles this disky post-merger is viewed edge-on with no obvious merger features that are identifiable by-eye. No reasonable astronomer would look at these images and decide with any certainty that this galaxy is a recent merger. 

Thus we find that even when merger features are present in a galaxy (indeed, all six merger identification methods can identify this as a merger in at least one viewing angle), there are some viewing angles for which mergers cannot be distinctly separated from non-mergers. In the case of the individual non-parametric morphology statistics, this merger is only detected in 79.9\%, 9.8\%, 52.4\%, and 81.7\% of viewing angles using asymmetry, outer asymmetry, shape asymmetry and the Gini-M$_{20}$ merger statistic, respectively. This improves to 92.5\% and 92.7\% when combined using the LDA and random forest methods. However, in the $\thicksim 8\%$ of viewing angles that the LDA and random forest methods cannot detect the merger, there is broad agreement with the human eye. Since galaxies are randomly oriented with respect to Earth, there is a $\thicksim 8\%$ chance of it being oriented such that the LDA or random forest (and perhaps even the human eye) could not differentiate it from a merger. This means there may be a fundamental upper limit to the completeness that can be achieved from the single viewing angle from Earth to any random galaxy in the Universe.

The results presented in Figure \ref{ViewAngExp} are limited to a single TNG100 galaxy. Generating 648 synthetic images and processing them with \texttt{statmorph} for all mergers drawn from TNG100 would be very computationally expensive. However, it is important to understand by how much different viewing angles affect merger detection on a broader statistical scale. For this reason, we take advantage of the four viewing angles which have already been generated and processed with \texttt{statmorph} for all 828 mergers and controls as presented in Sections \ref{Results-Ideal}-\ref{Results-props}. While four viewing angles for a single galaxy gives a much less detailed picture of how the morphology changes for a single galaxy, it does allow for a broad understanding of how ubiquitous and widespread the viewing angle limitation is across the entire merger sample. 

\begin{figure*}
    \centering
        \includegraphics[width=1\linewidth]{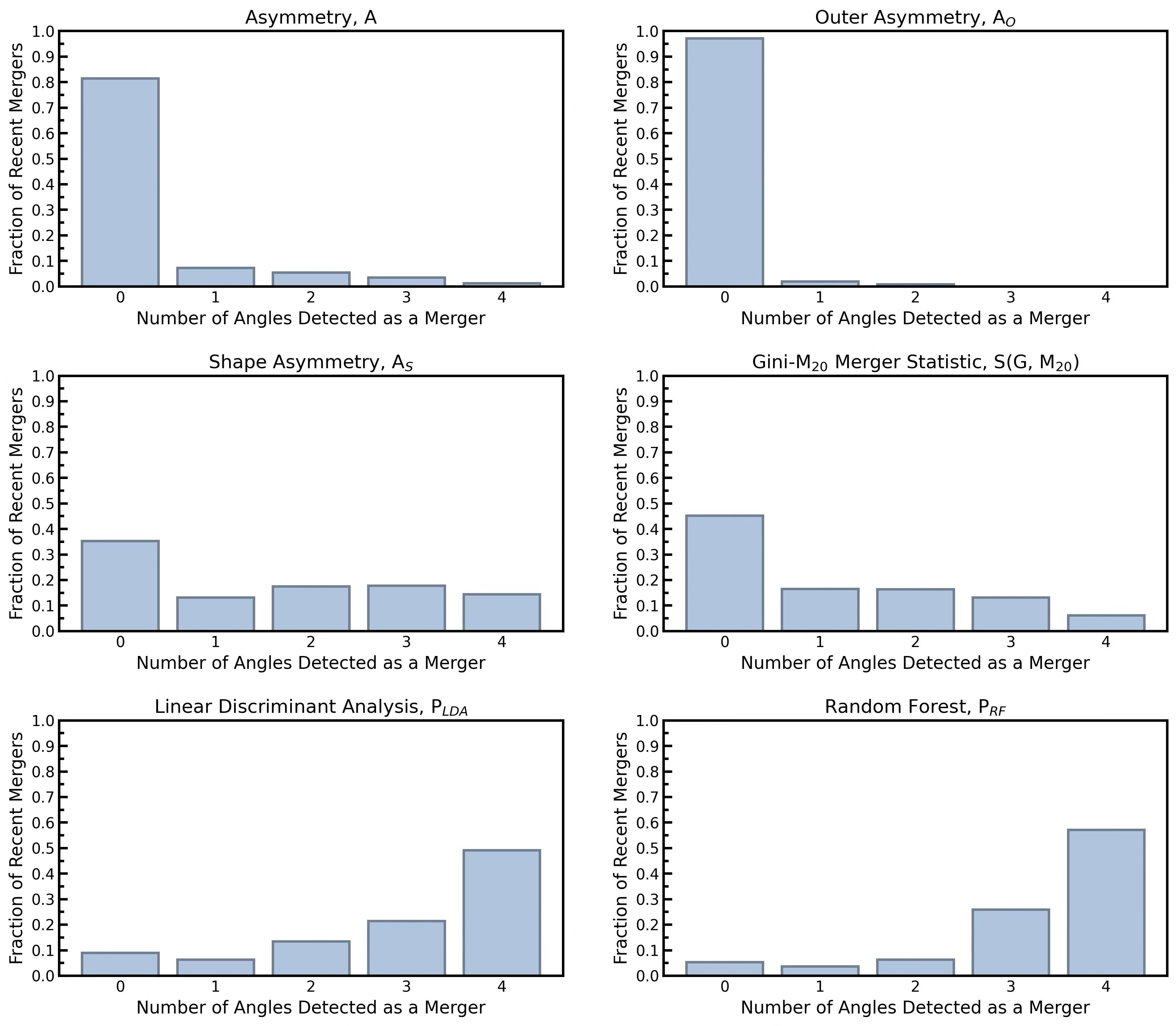}
        \caption{A histogram showing the fraction of the 424 recent post-mergers from TNG100 that were detected as a merger in 0, 1, 2, 3, or 4 of the four viewing angles generated for the entire merger sample (see Section \ref{SyntheticImages}). }
        \label{ViewAngExp-3}
    \end{figure*}

In Figure \ref{ViewAngExp-3}, we present the fraction of TNG100 post-mergers that are detected in 0, 1, 2, 3 or 4 of the 4 viewing angles used in Sections \ref{Results-Ideal} through \ref{Results-props}. For the entire sample of TNG100 mergers degraded to a depth of 26 mag arcsec$^{-2}$ and PSF FWHM of 0.75 arcsec, asymmetry, outer asymmetry, shape asymmetry and the Gini-M$_{20}$ merger statistic identify mergers on average in 9\%, 1\%, 41\% and 30\% of viewing angles, respectively. The LDA and random forest identify mergers in 74\% and 82\% of viewing angles, respectively. These values are very similar to the total completeness of merger sample at this image quality, as it is folding in all the same biases from the intrinsic galaxy properties. These mean values are substantially different from those reported for the single merger example in Figure \ref{ViewAngExp}, indicating the example selected was an outlier from the mean. The example galaxy was low mass and with a high gas fraction, which makes it more likely to be detected as a merger using asymmetry, for example (see Figure \ref{Completeness-fgas}).

The LDA and RF methods detect mergers in all four viewing angles 49\% and 57\% of the time, respectively. However, in the case of the individual non-parametric morphology statistics, a plurality of the mergers are detected in \emph{none} of the four viewing angles. In mergers where at least one viewing angle is detected as a merger, asymmetry, outer asymmetry, shape asymmetry and the Gini-M$_{20}$ merger statistic identify mergers on average in 47\%, 32\%, 63\% and 54\% of viewing angles. In other words, \textit{even when a galaxy has a morphological feature that an individual non-parametric statistic can use to identify it as a merger, due to orientation it will still be missed roughly half of the time.}







\section{Discussion}
\label{discussion}

\subsection{Context of Results Within Previous Works}

Reliably identifying galaxy mergers in an automated way is a challenge permeating throughout extragalactic research, as demonstrated by the efforts in the literature to understand the incidence rate of mergers in our Universe. Two approaches have been used to measure this incidence rate observationally. The first approach is to identify galaxies with physical separations and relative velocities indicative of an ongoing pair phase interaction \citep[e.g.][]{Man16, Mundy17, Duncan19}. In this way, every pair identified is one forthcoming merger. The second approach is that which is studied in detail in this work wherein recent merger events are identified by finding galaxies with post-merger features \citep[e.g.][]{Conselice09, Lotz11, Casteels14}. When differences in merger definitions are accounted for \citep[see][]{Lotz11, Husko22}, merger rates of the Universe measured using the former are generally higher than those measured using the latter \citep{Ren23}. 

The tension produced by these two methods is generally reconciled by the fact that observing post-merger features reliably is harder than detecting a separate companion for a number of reasons. Such reasoning has been informed by a wealth of previous works studying post-merger observability through high resolution (spatial and temporal) simulation suites of isolated mergers \citep[e.g.][]{Lotz08-mf, Lotz08-time, Lotz10-gas, Lotz10-mu, Wang12, Ji14, Amorisco15, Nevin19, VC22, McElroy22}. Firstly, isolated merger suites have shown that post-merger features are faint \citep{Wang12, Ji14, Amorisco15, VC22} and become harder to detect over time \citep{Lotz08-time, Lotz10-gas, Lotz10-mu, Nevin19, McElroy22}. Isolated merger suites have also been used to show that the detection of merger features is affected by the mass ratio of the progenitors \citep[e.g][]{Lotz10-mu, Ji14, Nevin19}, the gas fraction \citep[e.g.][]{Bell06, Lotz10-gas}, and orbital characteristics of the interaction \citep[e.g.][]{Naab03, Pawlik18, McElroy22}. While small merger suites tend to have better spatial and temporal resolution than cosmological simulations, the mergers are simulated with a select few initial conditions which do not adequately sample a diverse and realistic distribution of mass ratios, gas fractions, and orbital configurations. The discrete sampling of isolated merger suites limits our ability to translate these findings to a statistically robust understanding of critical effects on searching for post-mergers in real samples. 

Using cosmological simulations instead of idealized merger suites allows for an assessment of the reliability of merger identification in the context of realistic distributions of galaxy properties (stellar masses, mass ratios, and gas fractions) and orbital configurations. We can thus make a more realistic assessment of merger completeness, false positive rate, and purity. 

Several works have studied merger observability with non-parametic morphology statistics using cosmological simulations \citep{Bignone17, Snyder19, GO23}. For example, \citet{Bignone17} find that 45\% of major mergers are detected by $A>0.35$ in SDSS-like imaging. At slightly better image quality than that of SDSS we find only $\thicksim 10$\% of major mergers meet that same threshold (see central panel of Figure \ref{Completeness-mu}). We believe this discrepancy to be due to galaxies in the Illustris simulation being intrinsically more asymmetric than those in IllustrisTNG \citep[see Figure 6 of][]{RG19}. 

Our results are in close agreement with works that apply \texttt{statmorph} to cosmological simulations for the purposes of assessing merger observability \citep[e.g.][]{Snyder19, GO23, Rose23}. Using the higher resolution simulation IllustrisTNG50, \citet{GO23} find that using $A>0.25$ identifies 7.5\% of mergers in KiDS-realistic imaging. When accounting for the difference in threshold, we find that in our work, 8.5\% of mergers satisfy $A>0.25$ in similar image quality (FWHM = 0.75" and $\sigma_\text{sky} = 25$ mag arcsec $^{-2}$). \citet{GO23} also combine non-parametric morphology statistics together using a random forest model and achieve a completeness of 72\%. At the same image quality, we achieve a completeness of 85\%. A key difference between the random forests trained in this work is that shape asymmetry is the most important metric, whereas \citet{GO23} found asymmetry to be the most important. However, it is worth noting that the galaxy sample in \citet{GO23} contains more low-mass galaxies for which our results indicate asymmetry becomes a more useful merger indicator (see Figure \ref{Completeness-Mstar}). \citet{Snyder19} measure the non-parametric morphology statistics of realistic images from the original Illustris simulation and input these to a random forest for applications at higher redshift. At $z=0.5$, they achieve a completeness of $\thicksim 75$\% and find that Gini and M$_{20}$ are the most important features for their random forest model. This is loosely comparable to our results at our lowest image quality for which we achieved a completeness of 76\% and found Gini and M$_{20}$ to be the most important features for the random forest model. Similarly, \citet{Rose23} identify post-merger galaxies with a random forest trained on non-parametric morphology statistics derived from realistic imaging of TNG100 galaxies and achieve $\thicksim$ 60\% completeness out to $z = 4$. However, none of these previous works have tested the combined effect of image quality, stellar mass, or gas fraction on our ability to identify complete samples of mergers. In this work, we have brought together all of these elements in one place: an assessment of merger completeness with several merger identification methods and as a function of galaxy properties.

\subsection{Application of Results to Future Works}
\label{interpretation}

In this subsection, we offer suggestions based on our results for future works using non-parametric morphology statistics to identify mergers. First, recall that the grids of completeness, false positive rate, and purity in Figures \ref{ACompleteness}-\ref{RFCompleteness} can be used as a look-up table for the expected efficacy of those methods. Each cell represents a unique combination of depth and seeing in terms of $\sigma_\text{sky}$ in units of mag arcsec$^{-2}$ and PSF FWHM in units of arcseconds. However data from photometric surveys are rarely accompanied with noise statistics in terms of $\sigma_\text{sky}$. To facilitate translation between the sky noise used in this work and the depths reported by various photometric surveys, we have computed the depth of our synthetic imaging in terms of $\mu^\text{lim}_r$ (3$\sigma, $10" $\times$ 10") following the method described in \citet{Roman20} and adopting the notation from \citet{Martin22}, and 5$\sigma$ point source depths by injecting progressively brighter point sources into the noise until it is detected by Source Extractor Python \citep{Barbary2016} at 5$\sigma$. These values can be found in Table \ref{DepthConv}. Alternatively, $\sigma_\text{sky}$ of any imaging can be measured directly by simply taking the standard deviation of the background sky and converting it to units of mag arcsec$^{-2}$. Seeing is generally quoted in units of arcseconds. If the seeing is not known, it will have to be measured directly by fitting the point sources to a model PSF. Once the quality of imaging is known, \texttt{statmorph} can be run on the images and the completeness and false positive rate of each of the merger identification methods closest to the quality of imaging will be applicable to the out falling merger sample. However, if the merger threshold is altered, the completeness and false positive rate will no longer be applicable.

\begin{table}
\begin{center}
\begin{tabular}{|l|c|c|c|c|c|c|} 
 \hline
 $\sigma_\text{sky}$ & 23 & 24 & 25 & 26 & 27 & 28 \\ 
 \hline
 $\mu^\text{lim}_r$ (3$\sigma, $10" $\times$ 10") & 26.8 & 27.8 & 28.8 & 29.8 & 30.8  & 31.8\\
 \hline
 \hline
 5$\sigma$(FWHM $=1.50$") & 21.0 & 21.4 & 21.9 & 22.5 & 23.35 & 24.5 \\ 
 \hline
 5$\sigma$(FWHM $=1.25$") & 22.0 & 22.4 & 22.9 & 23.5 & 24.35 & 25.5 \\ 
 \hline
 5$\sigma$(FWHM $=1.00$") & 23.0 & 23.4 & 23.9 & 24.5 & 25.35 & 26.5 \\
 \hline
 5$\sigma$(FWHM $=0.75$") & 24.0 & 24.4 & 24.9 & 25.5 & 26.35 & 27.5 \\
 \hline
 5$\sigma$(FWHM $=0.50$") & 25.0 & 25.4 & 25.9 & 26.5 & 27.35 & 28.5 \\
 \hline
 5$\sigma$(FWHM $=0.25$") & 26.0 & 26.4 & 26.9 & 27.5 & 28.35 & 29.5 \\
 \hline

\end{tabular}
\end{center}
\caption{A conversion table between the measure of depth used in this work ($\sigma_\text{sky}$) and two other common measurements of depth, $\mu^\text{lim}_r$ (3$\sigma, $10" $\times$ 10") and 5$\sigma$ point source depth. $\mu^\text{lim}_r$ (3$\sigma, $10" $\times$ 10") is calculated following the method described in Appendix A of \citet{Roman20}. For a fixed amount of sky noise, 5$\sigma$ point source depth depends on the PSF, thus a full grid is required to convert $\sigma_\text{sky}$ to 5$\sigma$ point source depth. 5$\sigma$ point source depth also depends on the pixel scale of the imaging, although much less so than on the PSF FWHM.}
\label{DepthConv}
\end{table}

The purities we quote are only true for our \emph{balanced} dataset of equal mergers and non-mergers, and these values are subject to change depending on the relative frequency of mergers and non-mergers in a sample. For example, when measuring the incidence of mergers in the local Universe, the dataset will be lopsided with many more non-mergers than mergers which will decrease the purity of the merger sample, even if the completeness and false positive rate is constant. To demonstrate this point, the equation for purity (see equation \ref{PurityEq}) can be re-written in terms of the intrinsic merger fraction of a given sample ($f_m$), the number of galaxies in that sample ($N$), the completeness (COM) and false positive rate (FPR) of the merger identification method employed:

\begin{equation}
        \text{Purity} = \frac{\text{COM} \times f_m \times N}{\text{COM} \times f_m \times N + \text{FPR} \times (1 - f_m) \times N}.
        \label{PurityEq2}
    \end{equation}

In the case of a balanced dataset of mergers and non-mergers, $f_m = 0.5$, equation \ref{PurityEq2} reduces to equation \ref{PurityEq}, and all of the reported purities in Section \ref{allresults} are applicable. However, in the low-redshift Universe, the merger rate is generally much lower than this. Taking $f_m = 0.007$ \citep[see][]{Bickley21, GO23}, the purity from the random forest at the highest quality imaging becomes 2.7\% instead of 80\% in the balanced case.

The completeness and false positive rates for the merger detection methods used in this work can also be used to correct for the bias in any measured merger fraction enhancement relative to a control sample. Rearranging equation 4 from \citet{Lambrides21}, the absolute merger fraction enhancement in a given sample relative to a control sample, $\Delta f_m$, can be computed from the measured merger fraction enhancement, $\Delta \hat{f}_m$:

\begin{equation}
    \Delta f_m = \frac{\Delta \hat{f}_m}{\text{COM} - \text{FPR}}.
\end{equation}

\noindent However, our results in section \ref{Results-props} show that it is important that two samples are matched in redshift, mass, and gas fraction.

For future works using non-parametric morphology statistics to identify mergers, we recommend using a linear discriminant analysis or random forest model. The \textsc{scikit-learn} implementations of these algorithms are relatively straightforward and the additional amount of code required to train a model is minimal. A barrier to training an LDA or random forest for a new dataset is that pre-existing merger/non-merger labels for a subset of the data are required. However, with these classical machine learning techniques, it is not essential to have a large sample of training data. In Appendix \ref{Nlabels}, we show that a random forest supersedes the completeness of the individual non-parametric morphology statistics with only $\thicksim$100 truth labels. We do not, however, make any statement about any possible biases a model trained with so few labels may have. Nevertheless, it is not unprecedented to use visual classifications of a subset in order to train a model to classify the rest \citep[e.g.][]{Goulding18}. Deep learning techniques such as convolutional neural networks have a steeper learning curve and require more training data, but generally achieve higher completeness and purity than what we have presented in this work.

To facilitate the training of new models, our synthetic SKIRT images (with and without degradation to specific image qualities) are publicly available\footnote{https://www.canfar.net/storage/vault/list/AstroDataCitationDOI/
CISTI.CANFAR/23.0031/data/}. The images can be degraded to specific image qualities using the \texttt{RealSim} code from \citep{Bottrell19CNNReal}. Alternatively, our trained models can be used directly for merger detection. The LDA coefficients and how to use them can be found in Appendix \ref{LDAcoef}. The random forest decision trees can be loaded into Python from file and those files will be made public along with this paper\footnote{https://www.canfar.net/storage/vault/list/AstroDataCitationDOI/
CISTI.CANFAR/23.0031/data/TrainedModels/RF/}.







\section{Summary}
\label{Summary-TNG}

    Several non-parametric morphology statistics have been developed for the purpose of identifying mergers \citep[e.g.][]{Conselice03,Lotz04,Wen16,Pawlik16}. Previous works have demonstrated that the measured morphology statistics of galaxies are affected by the quality of imaging used \citep[e.g.][]{Lotz08-time, Pawlik16, Ferreira18, Thorp21, Deg23}. However, the individual relationships between PSF blurring, depth and the number of mergers that are identified using these statistics has not been extensively quantified. 
    
    In this work, we have tested how reliably these statistics can identify mergers using a sample of 424 known mergers, free from observational degradation, from IllustrisTNG100. The mergers are recent ($t_\text{post-merger} \lesssim 200$ Myr), low-redshift ($z_\text{sim} < 0.2$), and isolated ($r > 100$ kpc) and are matched in mass, gas fraction and redshift to a sample of 424 isolated non-mergers. In Section \ref{SyntheticImages}, we described how the simulation data is used to generate synthetic $r$-band images of varying image quality, incorporating realistic effects such as dust attenuation with the radiative transfer code SKIRT9 \citep{Baes20}. We generate synthetic images for the merger and control samples at a fixed redshift of $z=0.1$ and over a 6$\times$6 grid of PSF blurring and depth, with PSF FWHMs ranging from 0.25-1.5 arcsec and depths ranging from 23-28 mag arcsec$^{-2}$ (see Figure \ref{RealExample}). The image qualities therefore span from worse than SDSS to better than the 10-year LSST co-adds. The images were processed with \texttt{statmorph} \citep{RG19} and the measured non-parametric morphology statistics are used to quantify the completeness, false positive rate, and purity of the detected mergers. The main conclusions from this analysis are as follows:

    \begin{itemize}

        \item \textbf{In ideal imaging, free of atmospheric blurring and with minimal sky noise, the maximum completeness of the merger sample using non-parametric morphology statistics is 55.4\%.} Specifically, asymmetry, outer asymmetry, shape asymmetry and the Gini-M$_{20}$ merger statistic produce a completeness of 55.4\%, 36.9\%, 45.1\% and 49.4\% in the ideal imaging, respectively, at their default thresholds suggested in the literature (see Section \ref{Results-IQ} and Figure \ref{OneIQ-dists}).
        \smallskip

        \item \textbf{Combining non-parametric morphology statistics together using classical machine learning methods greatly improves upon the completeness and purity of individual non-parametric morphology statistics.} In the case of ideal imaging, the linear discriminant analysis and random forest methods trained on $A$, $A_O$, $A_S$, G, and M$_{20}$ achieved a completeness of 72.9\% and 86.4\%, respectively, with a false positive rate of only 15.0\% and 17.5\% (see Section \ref{Results-IQ} and Figure \ref{OneIQ-dists}).
        \smallskip
        
        \item \textbf{Higher quality imaging allows more mergers to be identified, increasing the completeness of the sample.} Over the range of image qualities and morphology statistics tested, the completeness of recovered mergers always improved from the lowest quality imaging to the highest quality imaging. However, asymmetry, outer asymmetry and the Gini-M$_{20}$ merger statistic also saw a proportional (or worse) increase in the false positive rate at higher image qualities. Thus, for these methods, increasing image quality did not allow for better distinction between mergers and non-mergers. For the shape asymmetry, LDA and random forest methods, completeness increased and false positive rate decreased in higher image qualities, leading to improved purity (see Section \ref{Results-IQ} and Figures \ref{ACompleteness}-\ref{GM20Completeness}).
        \smallskip
    



        \item \textbf{Even in high quality imaging, mergers identified with individual non-parametric morphology statistics are biased towards high mass ratio, high gas fraction, and low total stellar mass.} This is ubiquitous across all the individual non-parametric morphology statistics tested in this work. However, \textit{using the LDA or random forest methods significantly mitigates these merger detection biases} (see Section \ref{Results-props} and Figures \ref{Completeness-mu}-\ref{Completeness-fgas}).
        \smallskip

        \item \textbf{There is an upper limit on the completeness of recovered mergers caused by merger features not appearing in all viewing angles.} In an intermediate image quality, similar to that of many present day state-of-the-art photometric surveys, mergers were only recovered, on average, in 1-48\% of viewing angles, depending on the individual morphology statistic. In cases where a merger was detected in least one viewing angle, still only 32-63\% of viewing angles could be recovered as mergers by individual statistics. However, when using the LDA and random forest methods, mergers are detected in \emph{all four viewing angles} 49\% and 57\% of the time, respectively (see Section \ref{disc-ViewAngExp} and Figures \ref{ViewAngExp} and \ref{ViewAngExp-3}).
        \smallskip

    \end{itemize}

\section*{Acknowledgements}


We respectfully acknowledge the L\textschwa\textvbaraccent {k}$^{\rm w}$\textschwa\ng{}\textschwa n Peoples on whose traditional territory the University of Victoria stands and the Songhees, Esquimalt and $\underline{\text{W}}\acute{\text{S}}$ANE$\acute{\text{C}}$ peoples whose relationships with the land continue to this day. As we explore the shared sky, we acknowledge our responsibilities to honour those who were here before us, and their continuing relationships to these lands. We strive for respectful relationships and partnerships with all the peoples of these lands as we move forward together towards reconciliation and decolonization. 

We thank the anonymous reviewer of this work for their insight and detailed comments that helped strengthen our work. SW and SBM gratefully acknowledge the support from the Natural Sciences and Engineering Council of Canada (NSERC) as part of their graduate fellowship program. SLE and DRP gratefully acknowledge the receipt of NSERC Discovery Grants. Cette recherche a été financée par le Conseil de recherches en sciences naturelles et en génie du Canada (CRSNG). This research was enabled in part by the computing resources and support provided by the Digital Research Alliance of Canada (https://alliancecan.ca/en) and WestDRI (https://training.westdri.ca/).

\section*{Data Availability}

The LDA and random forest models trained on the image qualities tested in this work are available at www.canfar.net/citation/landing?doi=23.0031. Alternatively, new models can be trained using our synthetic SKIRT images (with and without degradation to specific image qualities) of TNG100 mergers and non-merger controls which are also available at the same online repository. The images can be degraded to any image qualities using the \texttt{RealSim} code which is available at https://github.com/cbottrell/RealSim. 

\bibliographystyle{mnras}
\bibliography{Bibliography}

\appendix
\newpage

\section{The Number of Galaxy Labels Needed to Train a LDA or Random Forest Model}
\label{Nlabels}

In Section \ref{allresults}, we find that when non-parametric morphology statistics are combined together with LDA or random forest methods, mergers and non-mergers can be better distinguished than when the statistics are used individually. We also found that the LDA and random methods could identify samples of mergers that were less biased with respect to stellar mass, mass ratio of progenitors, gas fraction, and orientation. However, training these methods for a new dataset requires pre-existing labels for that dataset which normally will not be available. In this section, we test how many labels are required to make training worthwhile.

    \begin{figure}
        \centering
        \includegraphics[width=1\linewidth]{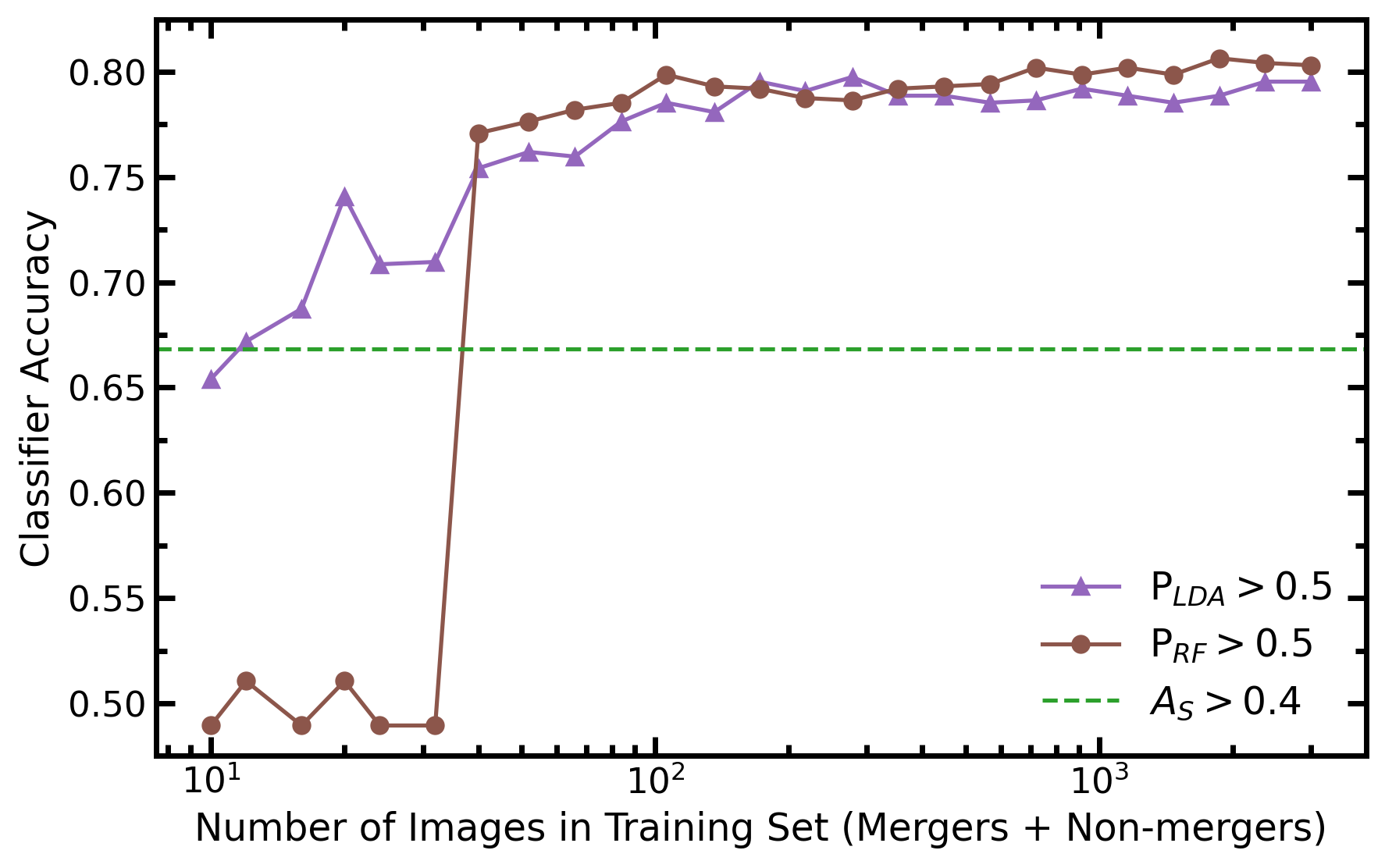}
        \caption{The LDA (purple triangles) and random forest (brown circles) accuracy on total test set when using a truncated training set. The number of images in the training set ranges from 10 to 3000 in logarithmic increments and a balanced number of mergers and non-mergers is enforced. The green dashed line is the accuracy achieved on the test set when identifying mergers with $A_S > 0.4$.}
        \label{RFAcc}
    \end{figure}

    In Figure \ref{RFAcc}, we present LDA and random forest accuracies as a function of the number of images included in the training set for one example image quality (FWHM = 0.75", $\sigma_\text{sky} = 26$ mag arcsec$^{-2}$). The images in these truncated training sets are drawn at random from the original training set from our main analysis, but required to produce a balanced training set of mergers and non-mergers. The number of images in the training set ranges from 10 to 3000 in logarithmic increments. For each size of training set, we compute the accuracy of the LDA and random forest classifier, as applied to the entirety of the test set from the main body of work. 
    
    When few images are used in the training of the models, the random forest achieves an accuracy of $\thicksim 50$\%, equivalent to a random classifier. Once the random forest has $\thicksim 20$ images to train on, the accuracy increases as a step function to greater than 75\%. The random forest accuracy increases with the number of images in the training set until after $\thicksim 100$ images the accuracy begins to plateau. The LDA classifier starts with an accuracy of 65\% and steadily increases to $\thicksim 80$\% once $\thicksim 200$ images are in the training set. We therefore recommend using at least 100 images to train a random forest and 200 to train a LDA model. Though, performance will continue to improve with more example images than this.

    The green dashed line in Figure \ref{RFAcc} is the accuracy achieved on the test set when identifying mergers with $A_S > 0.4$. The LDA and random forest achieve higher accuracies with fewer than 50 images. Therefore, even if there are only $\thicksim 50$ images in a test set (i.e. 25 correctly labelled mergers and 25 non-mergers), it is worthwhile to train an LDA or random forest model as it will achieve higher completeness than any individual non-parametric morphology statistic.

\section{LDA Coefficients}
\label{LDAcoef}

One benefit of LDA is that the model prediction can be computed using a linear combination of normalized inputs, with coefficients output by the LDA itself. The coefficients of the LDA models trained at each image quality be found in Table \ref{LDACoefTab}. Such a table allows anyone to use the LDA algorithms developed here to classify mergers and non-mergers in any dataset similar to one of the 36 image qualities presented here (see Section \ref{interpretation}). 

\begin{figure}
    \centering
    \includegraphics[width=1\linewidth]{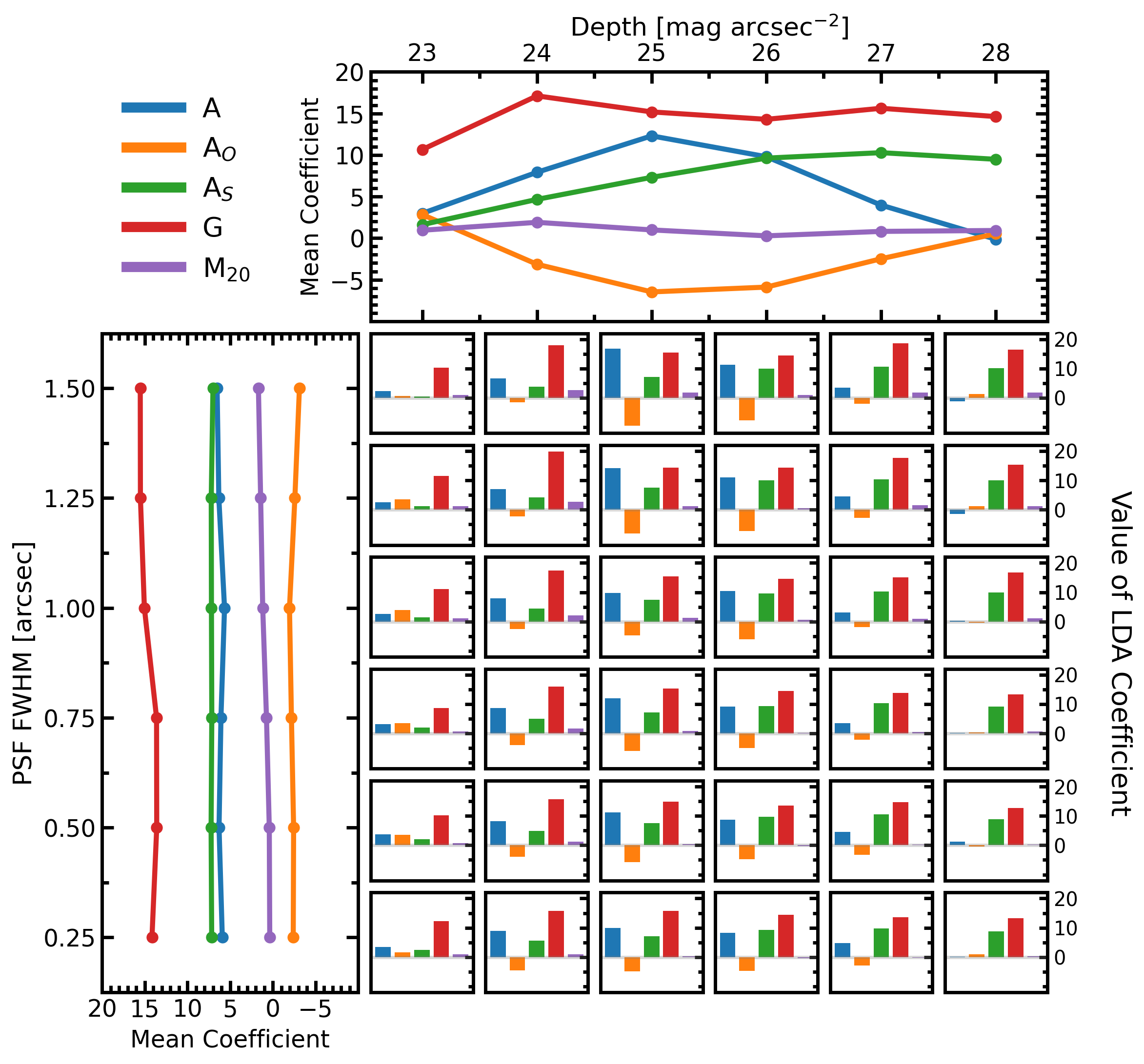}
        \caption{The LDA coefficients presented as a function of image quality. In the middle are 36 bar plots for each combination of depth and PSF FWHM. The colour of each bar represents the input statistic according to the legend in the top left of the figure and the height of the bars represents the associated coefficient for the optimized hyperplane separating mergers from non-mergers. The top panel shows the mean coefficient (at fixed depth) of each statistic as a function of depth and the left panel shows the mean coefficient (at fixed PSF FWHM) as a function of PSF FWHM}
        \label{LDACoefs}
    \end{figure}

In Figure \ref{LDACoefs}, we present the values of the LDA coefficients as a function of the image qualities tested in this work to unpack the LDA classification method in more detail. The figure follows the same structure as the LDA feature importance plot (Figure \ref{LDAImportance}), but now bar plots represent the strength of the LDA coefficients at each image quality. The top panel shows the mean coefficient at fixed depths for each input statistic as a function of depth. Likewise, the left panel shows the mean coefficient at fixed PSF FWHM. This left panel demonstrates that the LDA coefficients have no significant trends with seeing. However, several coefficients demonstrate more complicated trends with image depth (see top panel). At nearly all image qualities (34/36) Gini has the largest LDA coefficient and therefore correlates strongest with merger classification. In contrast, M$_{20}$ has the smallest coefficient at all image qualities indicating it is the least relevant statistic for LDA merger classification. Asymmetry has a small coefficient at low depths, increases to become a highly relevant statistic for merger classification at intermediate depths, only to decrease again to become almost entirely irrelevant at high depths. Outer asymmetry exhibits the inverse trend as asymmetry starting with a small positive coefficient, decreasing substantially to a strongly negative coefficient at intermediate depths indicating an anti-correlation with merger prediction, only to return to a near zero value at high depths. Shape asymmetry has a very low coefficient in shallow imaging which increases strongly with depth. At high depths, the LDA coefficients suggest that using only shape asymmetry and Gini statistics is the most efficient way to distinguish mergers from non-mergers.

    \begin{table}
    \begin{center}
    
    \begin{tabular}{|l|c|c|c|c|c|c|} 
    \hline
    $A$ & 23 & 24 & 25 & 26 & 27 & 28 \\ 
    \hline
1.5 & 2.28 & 6.71 & 16.72 & 11.27 & 3.40 & -1.19 \\
1.25 & 2.58 & 7.05 & 14.18 & 10.98 & 4.55 & -1.37 \\
1.0 & 2.58 & 7.88 &  9.82 & 10.35 & 3.13 &  0.22 \\
0.75 & 3.19 & 8.68 & 11.92 &  9.14 & 3.46 &  0.11 \\
0.5 & 3.76 & 8.28 & 11.28 &  8.72 & 4.51 &  1.28 \\
0.25 & 3.47 & 8.93 &  9.90 &  8.31 & 4.78 &  0.14 \\
    \hline
    \end{tabular}

    \smallskip
    
    \begin{tabular}{|l|c|c|c|c|c|c|} 
    \hline
    $A_O$ & 23 & 24 & 25 & 26 & 27 & 28 \\ 
    \hline
1.5 & 0.70 & -1.58 & -9.55 & -7.68 & -1.93 &  1.38 \\
1.25 & 3.57 & -2.32 & -8.02 & -7.19 & -2.71 &  1.20 \\
1.0 & 4.00 & -2.56 & -4.71 & -6.12 & -1.94 & -0.35 \\
0.75 & 3.55 & -3.96 & -5.95 & -4.91 & -2.17 &  0.38 \\
0.5 & 3.62 & -3.96 & -5.71 & -4.83 & -3.24 & -0.47 \\
0.25 & 1.59 & -4.50 & -4.89 & -4.65 & -2.90 &  1.02 \\
    \hline
    \end{tabular}

    \smallskip

    \begin{tabular}{|l|c|c|c|c|c|c|} 
    \hline
    $A_S$ & 23 & 24 & 25 & 26 & 27 & 28 \\ 
    \hline
1.5 & 0.50 & 3.86 & 7.14 & 9.91 & 10.55 & 10.15 \\
1.25 & 1.19 & 4.24 & 7.49 & 9.99 & 10.40 & 10.01 \\
1.0 & 1.47 & 4.44 & 7.50 & 9.61 & 10.22 &  9.93 \\
0.75 & 2.04 & 5.03 & 7.20 & 9.27 & 10.28 &  9.11 \\
0.5 & 2.03 & 4.85 & 7.54 & 9.66 & 10.48 &  8.86 \\
0.25 & 2.48 & 5.54 & 7.03 & 9.33 &  9.77 &  8.85 \\
    \hline
    \end{tabular}

    \smallskip

    \begin{tabular}{|l|c|c|c|c|c|c|} 
    \hline
    G & 23 & 24 & 25 & 26 & 27 & 28 \\ 
    \hline
1.5 & 10.4 & 17.9 & 15.4 & 14.4 & 18.7 & 16.4 \\
1.25 & 11.4 & 19.9 & 14.4 & 14.4 & 17.8 & 15.3 \\
1.0 & 11.1 & 17.5 & 15.3 & 14.5 & 15.2 & 16.8 \\
0.75 &  8.7 & 16.0 & 15.3 & 14.5 & 13.8 & 13.4 \\
0.5 & 10.2 & 15.7 & 14.9 & 13.6 & 14.7 & 12.6 \\
0.25 & 12.2 & 15.8 & 15.7 & 14.4 & 13.6 & 13.3 \\
    \hline
    \end{tabular}

    \smallskip

    \begin{tabular}{|l|c|c|c|c|c|c|} 
    \hline
    M$_{20}$ & 23 & 24 & 25 & 26 & 27 & 28 \\ 
    \hline
1.5 & 1.02 & 2.65 & 1.81 &  0.97 &  1.84 & 1.85 \\
1.25 & 1.21 & 2.79 & 1.19 &  0.60 &  1.51 & 1.28 \\
1.0 & 1.11 & 2.07 & 1.28 &  0.54 &  0.95 & 1.18 \\
0.75 & 0.70 & 1.67 & 0.85 &  0.15 &  0.52 & 0.60 \\
0.5 & 0.72 & 1.20 & 0.46 & -0.32 &  0.19 & 0.27 \\
0.25 & 0.92 & 1.04 & 0.35 & -0.29 & -0.15 & 0.33 \\
    \hline
    \end{tabular}

    \smallskip
    
    \end{center}
    \caption{LDA coefficients for the image qualities tested in this work. From top to bottom, the separated tables are for the $A$, $A_O$, $A_S$, G, and M$_{20}$ terms. In each individual table, the top row is the depth of the imaging in units of mag arcsec$^{-2}$ and the left row is the PSF FWHM in units of arcseconds.}
    \label{LDACoefTab}
    \end{table}

\section{Balance Point Thresholds}
\label{thresh}

Each of the merger identification methods have a default merger threshold suggested by the literature (see Section \ref{mergerid}). These thresholds are commonly used to classify galaxies as mergers and non-mergers and thus are used for assessing the efficacy of the methods for most of this work. However, previous studies have shown that non-parametric morphology statistics are sensitive to wavelength \citep{Kelvin12, Vika13, Haussler13,Baes20} and to different image qualities \citep{Lotz04, Moore06, Lisker08}. Thus, it would logically follow that the threshold separating mergers from non-mergers would also depend on these variables. 

In Section \ref{ROCResults}, we use the area under ROC curves of the merger identification methods to assess the potential of each method to differentiate between mergers and non-mergers, regardless of the merger threshold used. We report the balance point threshold of each merger identification method across the grid of image qualities used in this work in Table \ref{ThreshTab}. The balance point threshold is that which ensures $\text{COM} = 1 - \text{FPR}$. This threshold does not necessarily optimize for completeness, or purity, but -- as the name suggests -- strikes a balance between the two.

    \begin{table}
    \begin{center}
    
    \begin{tabular}{|l|c|c|c|c|c|c|} 
    \hline
    $A$ & 23 & 24 & 25 & 26 & 27 & 28 \\ 
    \hline
1.5 & -0.15 & -0.14 & -0.13 & -0.12 & -0.11 & -0.10 \\
1.25 & -0.07 & -0.06 & -0.05 & -0.03 & -0.02 &  0.00 \\
1.0 &  0.01 &  0.02 &  0.04 &  0.06 &  0.09 &  0.12 \\
0.75 &  0.09 &  0.10 &  0.12 &  0.15 &  0.18 &  0.21 \\
0.5 &  0.13 &  0.15 &  0.17 &  0.20 &  0.23 &  0.26 \\
0.25 &  0.16 &  0.17 &  0.20 &  0.22 &  0.25 &  0.29 \\
    \hline
    \end{tabular}

    \smallskip
    
    \begin{tabular}{|l|c|c|c|c|c|c|} 
    \hline
    $A_O$ & 23 & 24 & 25 & 26 & 27 & 28 \\ 
    \hline
1.5 & -0.13 & -0.12 & -0.12 & -0.11 & -0.10 & -0.09 \\
1.25 & -0.07 & -0.07 & -0.06 & -0.05 & -0.03 & -0.02 \\
1.0 &  0.00 &  0.01 &  0.03 &  0.05 &  0.06 &  0.08 \\
0.75 &  0.09 &  0.11 &  0.13 &  0.15 &  0.17 &  0.18 \\
0.5 &  0.18 &  0.20 &  0.22 &  0.24 &  0.25 &  0.27 \\
0.25 &  0.25 &  0.27 &  0.29 &  0.31 &  0.33 &  0.35 \\
    \hline
    \end{tabular}

    \smallskip

    \begin{tabular}{|l|c|c|c|c|c|c|} 
    \hline
    $A_S$ & 23 & 24 & 25 & 26 & 27 & 28 \\ 
    \hline
1.5 & 0.31 & 0.32 & 0.33 & 0.35 & 0.35 & 0.36 \\
1.25 & 0.29 & 0.29 & 0.30 & 0.31 & 0.32 & 0.32 \\
1.0 & 0.27 & 0.27 & 0.28 & 0.28 & 0.29 & 0.29 \\
0.75 & 0.25 & 0.25 & 0.26 & 0.26 & 0.26 & 0.27 \\
0.5 & 0.24 & 0.25 & 0.25 & 0.25 & 0.25 & 0.26 \\
0.25 & 0.25 & 0.25 & 0.25 & 0.25 & 0.26 & 0.26 \\
    \hline
    \end{tabular}

    \smallskip

    \begin{tabular}{|l|c|c|c|c|c|c|} 
    \hline
    G-M$_{20}$ & 23 & 24 & 25 & 26 & 27 & 28 \\ 
    \hline
1.5 & -0.07 & -0.07 & -0.06 & -0.06 & -0.06 & -0.05 \\
1.25 & -0.06 & -0.05 & -0.05 & -0.04 & -0.04 & -0.03 \\
1.0 & -0.05 & -0.05 & -0.04 & -0.04 & -0.03 & -0.03 \\
0.75 & -0.05 & -0.05 & -0.05 & -0.04 & -0.03 & -0.03 \\
0.5 & -0.05 & -0.05 & -0.05 & -0.04 & -0.03 & -0.03 \\
0.25 & -0.05 & -0.05 & -0.05 & -0.04 & -0.03 & -0.03 \\
    \hline
    \end{tabular}

    \smallskip

    \begin{tabular}{|l|c|c|c|c|c|c|} 
    \hline
    LDA & 23 & 24 & 25 & 26 & 27 & 28 \\ 
    \hline
1.5 & 0.50 & 0.49 & 0.48 & 0.48 & 0.50 & 0.48 \\
1.25 & 0.49 & 0.49 & 0.49 & 0.46 & 0.46 & 0.45 \\
1.0 & 0.48 & 0.50 & 0.48 & 0.47 & 0.44 & 0.47 \\
0.75 & 0.44 & 0.40 & 0.41 & 0.42 & 0.43 & 0.42 \\
0.5 & 0.44 & 0.45 & 0.41 & 0.44 & 0.43 & 0.43 \\
0.25 & 0.39 & 0.39 & 0.43 & 0.43 & 0.41 & 0.46 \\
    \hline
    \end{tabular}

    \smallskip

    \begin{tabular}{|l|c|c|c|c|c|c|} 
    \hline
    RF & 23 & 24 & 25 & 26 & 27 & 28 \\ 
    \hline
1.5 & 0.54 & 0.52 & 0.52 & 0.52 & 0.52 & 0.51 \\
1.25 & 0.55 & 0.51 & 0.52 & 0.49 & 0.52 & 0.49 \\
1.0 & 0.56 & 0.56 & 0.59 & 0.54 & 0.53 & 0.55 \\
0.75 & 0.57 & 0.51 & 0.53 & 0.52 & 0.54 & 0.55 \\
0.5 & 0.55 & 0.53 & 0.50 & 0.55 & 0.53 & 0.50 \\
0.25 & 0.52 & 0.52 & 0.51 & 0.55 & 0.50 & 0.60 \\
    \hline
    \end{tabular}

    \smallskip
    
    \end{center}
    \caption{Balance point thresholds for the image qualities tested in this work. From top to bottom, the separated tables are for the $A$, $A_O$, $A_S$, S(G, M$_{20}$), LDA, and random forest methods. In each individual table, the top row is the depth of the imaging in units of mag arcsec$^{-2}$ and the left row is the PSF FWHM in units of arcseconds.}
    \label{ThreshTab}
    \end{table}

\section{Asymmetry (Difference of Squares)}
\label{ARobust}

    \citet{Thorp21} demonstrates that the method of correcting for background noise in the asymmetry measurement used by \texttt{statmorph} (and indeed, in most asymmetry calculations), over-corrects for the noise, introducing a bias towards lower asymmetry values in noisy/shallow imaging. Recent works by \citet{Deg23} and \citet{Yu23} introduce new methods of measuring asymmetry which allow for sky noise to be more accurately accounted for, which has shown to decrease the bias introduced by noisy data. Here, we implement into the \texttt{statmorph} framework the \citet{Deg23} description of asymmetry which follows the same principle calculation as standard asymmetry but computes the sum as a difference of squares rather than as an absolute difference:

    \begin{equation}
        A_{sq} = \left(\frac{\Sigma (I_0 - I_{180})^2 - B_{sq}}{\Sigma (I_0 + I_{180})^2 - B_{sq}}\right)^{1/2}.
        \label{Asq_eq}
    \end{equation}

    This difference of squares method also differs in the way the asymmetry of the background is removed, subtracting the background term $B_{sq}$ from both the numerator and the denominator. $B_{sq}$ is measured by computing the difference of squares sum between the skybox and its 180$^\circ$-rotated counterpart:

    \begin{equation}
        B_{sq} = \Sigma (B_0 - B_{180})^2.
        \label{Bsq}
    \end{equation}

    It is also worth noting that \citet{Deg23} suggest that $B_{sq} \approx 2N\sigma^2$ where N is the number of pixels over which the $A_{sq}$ sum is computed and $\sigma$ is the standard deviation of the sky noise. We found that for the images used in this work, this approximation produced nearly identical results as Equation \ref{Bsq}. However, for imaging with worse pixel scale resolution, approaching that of the SDSS (0.396 arcsec / pix), the approximation was observed to break down. 

    \begin{figure}
        \centering
        \includegraphics[width=1\linewidth]{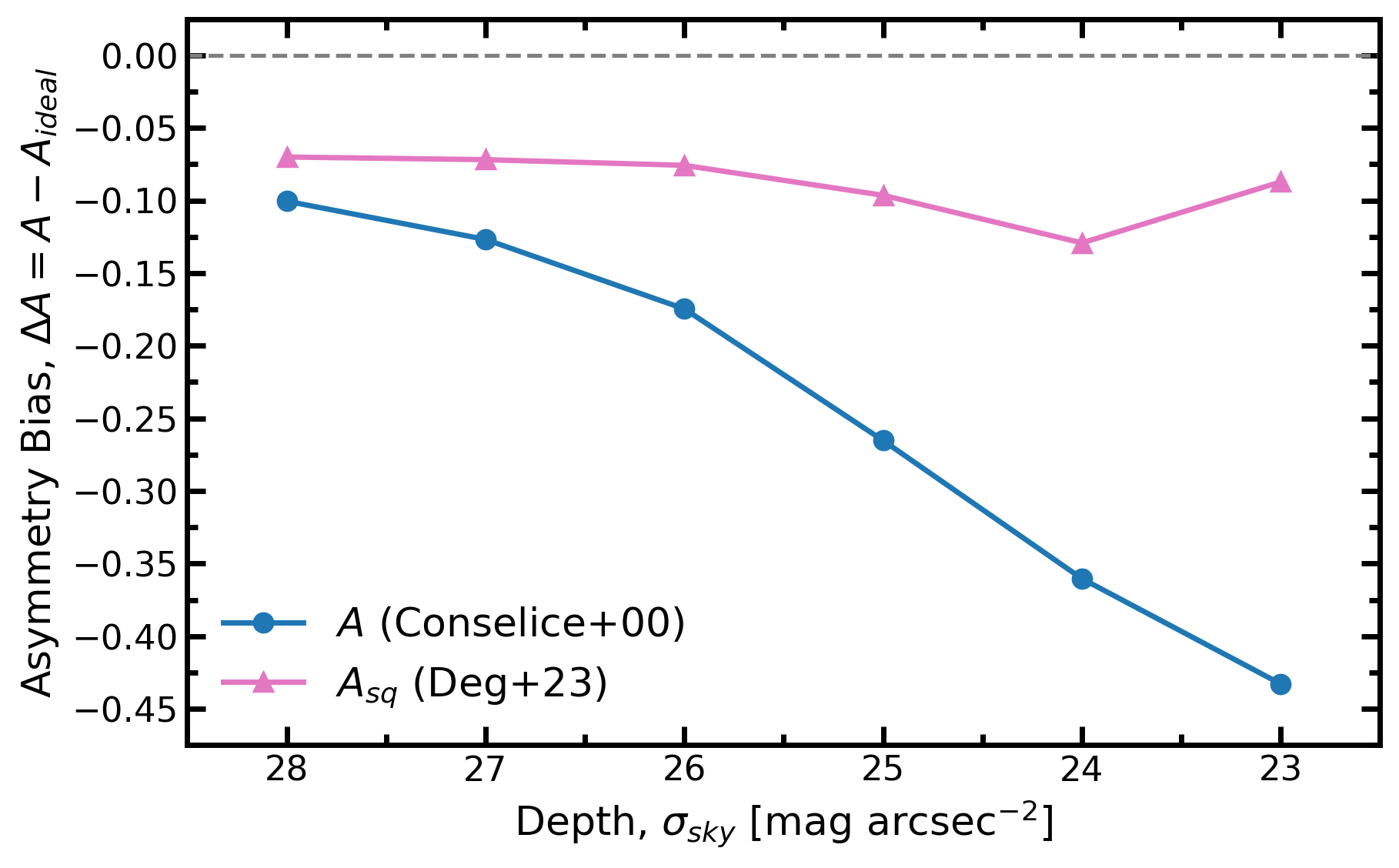}
        \caption{The observed asymmetry bias for the standard asymmetry definition \citep{Conselice00}, indicated by blue circles, and the difference of squares implementation of asymmetry \citep{Deg23}, indicated as pink triangles, as a function of increasing sky noise at a fixed PSF FWHM (0.75"). The asymmetry bias ($\Delta A$) is computed by taking the mean difference between the asymmetry measurement for a synthetic image of a isolated control galaxy with realistic sky noise and PSF ($A$) and the asymmetry measurement for the same image without any degradation ($A_\text{ideal}$).}
        \label{deltaA}
    \end{figure}
    
    In Figure \ref{deltaA}, we present the observed asymmetry bias for the standard asymmetry definition \citep{Conselice00} and the difference of squares implementation of asymmetry \citep{Deg23} as a function of increasing sky noise at a fixed PSF FWHM (0.75"). The asymmetry bias ($\Delta A$) is computed by taking the mean difference between the asymmetry measurement for a synthetic image of a isolated control galaxy with realistic sky noise and PSF ($A$) and the asymmetry measurement for the same image without any degradation ($A_\text{ideal}$). The blue curve demonstrates that in deep imaging, the asymmetry measurement is lower than the ideal asymmetry by $\thicksim 0.1$, and that as the sky noise increases, the asymmetry bias worsens substantially. In the most shallow imaging tested, the standard asymmetry measurement found control galaxies to have lower asymmetries than in the ideal imaging by $\thicksim 0.45$, on average. The pink curve demonstrates that at all image qualities, the \citet{Deg23} asymmetry definition exhibits a smaller bias due to sky noise. In deep imaging, it is only $\thicksim 0.075$ lower than in the ideal case and reaches a maximum average bias of only $\thicksim 0.13$ at a depth of $\sigma_\text{sky} = 24$ mag arcsec$^{-2}$. Increasing or decreasing the fixed PSF FWHM does not change the observed trends with depth and shifts the trend lines shown in Figure \ref{deltaA} vertically.
    

    Having established that the \citet{Deg23} asymmetry definition is less biased in noisy imaging, we now explore its potential for merger identification. Since this statistic has never been used to identify mergers previously, there is no suggested threshold from the literature above which mergers are expected to be found. To compare directly with the standard asymmetry results (see Section \ref{Aresults}), we take the merger threshold to be $A_{sq} > 0.35$. 
    
\begin{figure*}
    \centering
    \includegraphics[width=1\linewidth]{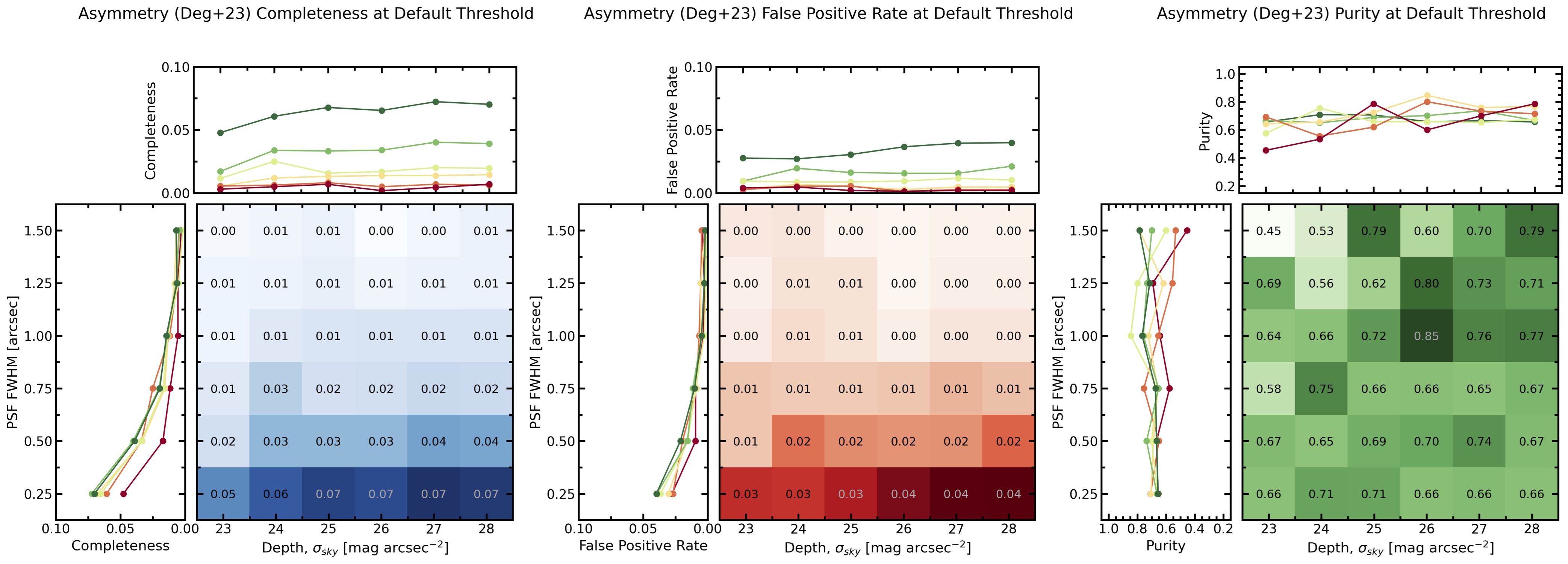}
        \caption{The completeness, false positive rate, and purity of the merger sample, as computed using the \citet{Deg23} asymmetry definition with the threshold $A_{sq}>0.35$. This figure is composed of three subfigures, one for each the completeness, false positive rate, and purity. Each subfigure contains three panels. The bottom right panel is the most important; the blue/red/green colour gradient shows qualitatively the trend of completeness/false positive rate/purity as a function of both depth and PSF blurring, with specific completeness values at each image quality reported in the corresponding cell. The top panel shows the relationship between completeness/false positive rate/purity and depth, with each line representing a different PSF FWHM. The lines are coloured from red to yellow to green with red indicating the worst image quality (highest PSF FWHM) and green indicating best image quality (lowest PSF FWHM). The left panel shows the relationship between completeness and resolution, with each line representing a different depth. These lines are also coloured from red to yellow to green with red indicating the worst image quality (lowest depth) and green indicating best image quality (high depth).}
        \label{AsqCompleteness}
    \end{figure*}

    In Figure \ref{AsqCompleteness}, we present the completeness of the merger sample recovered using the threshold $A_{sq} > 0.35$ as a function of depth and PSF blurring. The top panels of the completeness and false positive rate subfigures show that completeness and false positive rates are much more stable with depth (no significant trend) than in the case of standard asymmetry. However, the left panels show that $A_{sq}$ does not remove the dependence of  completeness and false positive rate on resolution. Broadly speaking, $A_{sq}$ achieves higher completeness (and false positive rate) than $A$ in shallow imaging, and lower completeness (and false positive rate) than $A$ in deep imaging. The performance of $A_{sq}$ at all depths is similar to that of $A$ at an intermediate depth ($\sigma_\text{sky} = 25$ mag arcsec$^{-2}$). We have therefore demonstrated that the \citet{Deg23} asymmetry definition improves the consistency of asymmetry across image depths. However, the \citet{Deg23} asymmetry definition is still consistently poor for the purposes of merger identification.
    

\bsp	
\label{lastpage}
\end{document}